\documentclass[%
reprint,
superscriptaddress,
 amsmath,amssymb,
 aps,
prb,
nobalancelastpage,
]{revtex4-2}

\usepackage{amsmath}
\usepackage{graphicx}
\usepackage{dcolumn}
\usepackage{bm}
\usepackage{hyperref}
\usepackage{url}
\usepackage{xcolor}

\hypersetup{colorlinks=true,allcolors=blue}

\newcommand{\Sr}{\texorpdfstring{Sr$_{2}$RuO$_{4}$}{Sr2RuO4}}

\begin{document}

\preprint{APS/123-QED}

\title{Finite-momentum and field-induced pairings in orbital-singlet spin-triplet superconductors}

\author{Jonathan Clepkens}
\affiliation{Department of Physics and Center for Quantum Materials,\\ University of Toronto, 60 St. George St., Toronto, Ontario, M5S 1A7, Canada}
\author{Hae-Young Kee}
 \email{hy.kee@utoronto.ca}
\affiliation{Department of Physics and Center for Quantum Materials,\\ University of Toronto, 60 St. George St., Toronto, Ontario, M5S 1A7, Canada}
\affiliation{Canadian Institute for Advanced Research, Toronto, Ontario, M5G 1Z8, Canada}


\begin{abstract}
Finite-momentum pairing in a Pauli-limited spin-singlet superconductor arises from the pair-breaking effects of an external Zeeman field, a mechanism which is not applicable in odd-parity spin-triplet superconductors. However, in multiorbital systems, the relevant bands originating from different orbitals are usually separated in momentum space, implying that orbital-singlet pairing is a natural candidate for a finite-momentum pairing state. We show that finite-momentum pairing arises in even-parity orbital-singlet spin-triplet superconductors via the combination of orbitally-nontrivial kinetic terms and Hund's coupling. The finite-momentum pairing is then suppressed with an increasing spin-orbit coupling, stabilizing a uniform pseudospin-singlet pairing. We also examine the effects of the magnetic field and find field-induced superconductivity at large fields. We apply these findings to the multiorbital superconductor with spin-orbit coupling, Sr$_{2}$RuO$_{4}$ and show that a finite-momentum pseudospin-singlet state appears between the uniform pairing and normal states. Future directions of inquiry relating to our findings are also discussed.
\end{abstract}

\maketitle

\section{\label{intro}Introduction}

Superconductivity is conventionally understood as the condensation of electron pairs near the Fermi energy with momentum $\textbf{k}$ and $-\textbf{k}$, i.e., a zero center-of-mass momentum, referred to as uniform pairing. 
A term in the Hamiltonian that breaks the degeneracy of the paired electrons generally suppresses the pairing. For instance, in the case of a spin-singlet superconductor (SC), the Zeeman field acts as a pair breaker and destroys the superconductivity. However, before the normal state is reached,  a finite center-of-mass momentum, $\textbf{q}$ (finite-q) pairing known as the Fulde-Ferrell-Larkin-Ovchinnikov (FFLO) state \cite{Fulde1964, Larkin1964} is found between the uniform pairing and normal states as the Zeeman field increases. Unlike the spin-singlet pairing, a finite-q pairing in a spin triplet may require a different mechanism, because the Zeeman field acting on the triplet pairs will suppress only the component of the $d$ vector parallel to the field. If the $d$ vector is not pinned along a certain direction, it can rotate under the field and thus an FFLO state is not expected.

A natural question is what the potential mechanisms for finite-q pairing in spin-triplet SCs are. Finite-q pairing that occurs without the explicit breaking of time-reversal symmetry is often referred to as a pair-density wave (PDW) \cite{agterberg2020}, and has been conjectured to be present in a wide variety of correlated materials of interest \cite{hamidian_Nat2016,chen_Nat2021, Liu_Science2021, gu_Nat2023, aishwarya_Nat2023, liu_Nature2023, zhao_Nature2023}. The putative spin-triplet SC, UTe$_{2}$, is one of the materials which has shown evidence for PDW order \cite{gu_Nat2023, aishwarya_Nat2023}. The Fe-based SCs and Sr$_{2}$RuO$_{4}$ have displayed signatures of finite-q pairing \cite{Cho_PRB2011,Tarantini_PRB2011,Kasahara_PRL2020,liu_Nature2023, zhao_Nature2023, Kinjo_Science2022} and an interorbital spin-triplet order parameter has been argued as a plausible option for both of them \cite{SpalekPRB2001,Dai2008PRL,Puetter2012EPL,Hoshino2015PRL,Hoshino2016PRB,vafek2017hund,Gingras2019PRL,lindquist2019distinct,Coleman_PRL2020, Suh2019, clepkens2021higher}. While the possibility of finite-q spin-triplet pairing 
remains a relevant topic for these correlated SCs, various theoretical models have focused on spin-singlet finite-q pairings \cite{Berg_PRB2009, jiang_2023, Jian_PRL2015, Jiang_PRL2023, chen_SCP2023, Han_PRB2022, chakraborty_2023, Kang_PRB2011, Lee_PRL2007, SotoGarrido_PRB2015, Florian_PRL2011, Wang_PRL2015, Wu_PRL2023, Fradkin_PRB2014, ticea_2024, Yerin_JPCM2023}, and finite-q pairing involving spin-triplet SCs remains relatively less explored.

Here, we explore a mechanism for generating finite-q pairing in multiorbital systems. In these systems, various types of superconducting order parameters exist. Notably, the even-parity orbital-singlet spin-triplet (OSST) pairing, which implies an antisymmetric wavefunction under the exchange of two orbitals, has been proposed for multiorbital SCs such as Sr$_{2}$RuO$_{4}$ and Fe-based SCs \cite{SpalekPRB2001,Dai2008PRL,Puetter2012EPL,Hoshino2015PRL,Hoshino2016PRB,vafek2017hund,Gingras2019PRL,lindquist2019distinct,Coleman_PRL2020, Suh2019, clepkens2021higher}. 
As orbital degeneracy is essential for uniform orbital-singlet pairing, we anticipate that terms that break the orbital degeneracy, serving as pair breakers, would hinder the uniform pairing
and finite-q pairing may emerge before reaching the normal state.

To illustrate this concept, we first examine a simple two-orbital model featuring Hund's coupling, resulting in an attractive interaction in the OSST channel within mean-field (MF) theory. In this model, the orbital-dependent potential, known as orbital polarization, along with the interorbital hopping termed orbital hybridization, gives rise to finite-q pairing. We also study the effects of spin-orbit coupling (SOC) and a Zeeman field, and find the appearance of a Fulde-Ferrell (FF) state for small fields and large SOC.
For large fields, we find a field-induced finite-q and uniform pairing phase, which survives for a small range of SOC values.

We then investigate the application to a SC involving three $t_{2g}$ orbitals, such as Sr$_{2}$RuO$_{4}$. Both the SOC and Hund's interaction have been recognized as playing an important role in understanding the superconductivity in Sr$_{2}$RuO$_{4}$ \cite{Ng2000EPL,Haverkort2008PRL,Kim2018PRL,Tamai2019PRX, Veenstra2014PRL, mravlje2011PRL, Georges2013ARCMP}. While the fundamental physics does not change with three orbitals, the differences introduced by going beyond two orbitals are discussed. The OSST finite-q pairing with a wavevector along the $x$-axis emerges between the uniform pairing and normal states with the introduction of the in-plane magnetic field.

The paper is organized as follows. Since the finite-q pairing requires a perturbation that hinders the uniform pairing, we discuss possible pair breakers in the next section.  A generic microscopic model for two orbitals, including the kinetic terms and SOC, along with the MF pairing terms is introduced in Sec.~\hyperref[Three]{III}. The phase diagram for OSST pairing as a function of both the orbital polarization and SOC is obtained. We also obtain the phase diagram for uniform and finite-q phases as a function of magnetic field and SOC in Sec.~\hyperref[Four]{IV}. In Sec.~\hyperref[Five]{V}, we use a three-orbital model to discuss the potential applications of our findings to Sr$_{2}$RuO$_{4}$, and we discuss further directions of inquiry in the last section.

\section{\label{Two}Pair breakers for uniform OSST pairing}
In this section, we provide an intuitive discussion on generating finite-q OSST pairing before proceeding to the MF calculations. Considering that the Zeeman field acts as a pair breaker for the spin singlet by polarizing the spins, any term disrupting the degeneracy of orbitals serves as a pair breaker for the OSST and may induce finite-q pairing before reaching the normal state.

To formulate such effects, let us consider a two-orbital MF Hamiltonian describing the uniform superconducting state. We introduce the basis, $\psi_{\textbf{k}}^{\dagger} = (c^{a\dagger}_{\textbf{k}\uparrow}, c^{a\dagger}_{\textbf{k}\downarrow}, c^{b\dagger}_{\textbf{k}\uparrow}, c^{b\dagger}_{\textbf{k}\downarrow})$, which consists of creation operators for an electron in one of the two orbitals $a,b$ with spin $\sigma=\uparrow,\downarrow$. Using the Nambu spinor, $\Psi_{\textbf{k}}^{\dagger} = (\psi_{\textbf{k}}^{\dagger}, \psi_{-\textbf{k}}^{T})$, the uniform MF Hamiltonian is,
\begin{equation}
    H_{\text{MF}} = \sum_{\textbf{k}}\Psi_{\textbf{k}}^{\dagger}\begin{pmatrix}
   H(\textbf{k}) & \Delta(\textbf{k}) \\[6pt]
   \Delta^{\dagger}(\textbf{k}) & -H^{*}(-\textbf{k}) \\
 \end{pmatrix}\Psi_{\textbf{k}},
\end{equation}
where the normal-state Hamiltonian $H(\textbf{k})=\sum_{\gamma\gamma^{\prime}}H^{\gamma\gamma^{\prime}}(\textbf{k})\tau_{\gamma}\sigma_{\gamma^{\prime}}$, and the gap matrix $\Delta(\textbf{k})=\sum_{\gamma\gamma^{\prime}}\Delta^{\gamma\gamma^{\prime}}(\textbf{k})\tau_{\gamma}\sigma_{\gamma^{\prime}}(i\sigma_{2})$ are written in terms of the Pauli and identity matrices, $\tau_{\gamma} (\sigma_{\gamma^{\prime}})~(\gamma,\gamma^{\prime} = 0,...3)$ in orbital(spin) space. The terms in $H(\textbf{k})$ given by the sum over pairs of indices $(\gamma,\gamma^{\prime})$ are determined by the symmetry and other microscopic considerations, while the terms in the sum for the gap matrix are determined by the type of pairing under consideration.

\begin{table}[t!]
\centering
\begin{tabular}{|c c | c c|} 
 \hline
  SC gap &  & Pair breaker &  \\ [0.5ex] 
 \hline\hline
 OTSS: & $\Delta^{00}\tau_{0}(i\sigma_{2})$ &  Zeeman field: & -$\tau_{0}(\boldsymbol{h}\cdot\boldsymbol{\sigma})$ \\ [1ex]
 \hline
OSST: & $i\tau_{2} \boldsymbol{\Delta}_{\text{OSST}}\cdot\boldsymbol{\sigma}(i\sigma_{2})$ & Orbital & \vspace{-5pt}   \\ & & polarization: & $h^{z}_{P}(\textbf{k})\tau_{3}\sigma_{0}$ \\[1ex] & & Orbital &   \vspace{-5pt}\\ & & hybridization: &  $h^{x}_{P}(\textbf{k})\tau_{1}\sigma_{0}$  \\[1ex]
 \hline
\end{tabular}
\caption{Terms in the normal-state Hamiltonian, $H^{\gamma\gamma^{\prime}}(\textbf{k})\tau_{\gamma}\sigma_{\gamma^{\prime}}$, which have a pair-breaking effect on a gap matrix, $ \Delta^{\gamma\gamma^{\prime}}(\textbf{k})\tau_{\gamma}\sigma_{\gamma^{\prime}}(i\sigma_{2})$. Any term which breaks the time-reversal symmetry is a pair-breaker for the conventional OTSS pairing,  $\Delta^{00}$, with the Zeeman field, $\boldsymbol{h}\equiv(H^{01}, H^{02}, H^{03})$, listed in the second row. The third and fourth row are pair breakers for OSST pairing, $\boldsymbol{\Delta}_{\text{OSST}} \equiv (\Delta^{21}, \Delta^{22}, \Delta^{23})$. The OSST pair breakers are written as $(\boldsymbol{h}_{P}(\textbf{k})\cdot\boldsymbol{\tau})\sigma_{0}$
where $\boldsymbol{h}_{P}(\textbf{k})\equiv(H^{10}(\textbf{k}), 0, H^{30}(\textbf{k}))$. Since the second component breaks time-reversal symmetry, it is neglected.}
\label{t1}
\end{table}

To understand the pair breakers of the OSST, let us recall the conventional spin singlet case first. The conventional orbital-triplet spin-singlet (OTSS) pairing is $\Delta^{00}(\textbf{k})\tau_{0}i\sigma_{2}$, and the Zeeman term with arbitrary direction of the magnetic field, $-\tau_{0}(\boldsymbol{h}\cdot\boldsymbol{\sigma})$ where $\boldsymbol{h}\equiv (H^{01}, H^{02}, H^{03})$, suppresses the pairing by breaking the spin degeneracy. The OSST pairing, on the other hand, is represented by $ i\tau_{2}\boldsymbol{\Delta}_{\text{OSST}}\cdot\boldsymbol{\sigma}(i\sigma_{2})$, with $\boldsymbol{\Delta}_{\text{OSST}}\equiv(\Delta^{21}, \Delta^{22}, \Delta^{23})$, which can become an attractive channel within MF theory due to the Hund's coupling \cite{klejnberg1999hund, SpalekPRB2001, HanPRB2004, Dai2008PRL, Puetter2012EPL, vafek2017hund}. In analogy to the Zeeman term $-\tau_{0}(\boldsymbol{h}\cdot\boldsymbol{\sigma})$ being a pair breaker for the OTSS,
the pair-breaking terms for the OSST pairing can be written as, $(\boldsymbol{h}_{P}(\textbf{k})\cdot\boldsymbol{\tau})\sigma_{0}$
where $\boldsymbol{h}_{P}(\textbf{k})\equiv(H^{10}(\textbf{k}), H^{20}(\textbf{k}), H^{30}(\textbf{k}))$. These terms suppress the OSST pairing by breaking the orbital degeneracy. The first component, $h^{x}_{P}(\textbf{k})\tau_{1}\sigma_{0}$, corresponds to a spin-independent hybridization between $a,b$ orbitals, for example via interorbital hopping. The second component corresponding to the term $h^{y}_{P}(\textbf{k})\tau_{2}\sigma_{0}$ is neglected to preserve time-reversal symmetry.  The third component, $h^{z}_{P}(\textbf{k})\tau_{3}\sigma_{0}$, corresponds to an orbital-dependent potential or dispersion and is termed `orbital polarization', which may arise from the two orbitals having different hopping integrals for instance.

The OSST pairing along with the respective terms in the Hamiltonian that act as pair breakers are summarized in Table~\ref{t1}. For comparison, the conventional OTSS is also listed.  Note that the Zeeman field does not have a pair-breaking effect on OSST pairing in the absence of a pinning of the $d$ vector (for example, via SOC), as discussed previously. 
Another example that generates the orbital polarization is the nematic order,  i.e., a spontaneous rotational symmetry breaking in the density of the two orbitals, such as $\langle{n_{yz}-n_{xz}}\rangle$ for $(d_{yz},d_{xz})$ orbitals on a square lattice. Alternatively, $h_{P}^{z}$ can have a $\textbf{k}$-dependence due to an anisotropy in the dispersions of the orbitals. In the next section we discuss a concrete example to make the origin of such terms clear.

\section{\label{Three}finite-q OSST pairing in zero field}

To elucidate the mechanism for generating finite-q pairing from the uniform OSST pairing, we will now consider the case of two orbitals $a,b$ related by $C_{4}$, such as $(d_{yz}, d_{xz})$ orbitals. Later on we will comment on more general cases. The orbital dispersions are, $\xi^{a/b}_{\textbf{k}} = -2t_{1}\cos{k_{y/x}} -2t_{2}\cos{k_{x/y}} -\mu$, with $t_{1}$ and $t_{2}$ the nearest neighbor (NN) hoppings along the $y(x)$- and $x(y)$-axis for $a(b)$ orbitals respectively. The orbital hybridization originating from interorbital hopping is $t_{\textbf{k}} = -4t_{ab}\sin{k_x}\sin{k_y}$. The minimal normal-state Hamiltonian respecting time-reversal and inversion is then,
\begin{equation}
    H(\textbf{k}) = \frac{\xi^{+}_{\textbf{k}}}{2}\tau_{0}\sigma_{0} + \frac{\xi^{-}_{\textbf{k}}}{2}\tau_{3}\sigma_{0} + t_{\textbf{k}}\tau_{1}\sigma_{0}.
\label{kinetic}
\end{equation}
We have defined $\xi_{\textbf{k}}^{\pm}=\xi^{a}_{\textbf{k}}\pm\xi^{b}_{\textbf{k}}$, and thus $\xi^-_{\textbf{k}}$ plays the role of the ${\textbf{k}}$-dependent orbital polarization, equivalent to $h_{P}^{z}(\textbf{k})$ listed in Table~\ref{t1}, while $t_{\textbf{k}}$ is equivalent to $h_{P}^{x}(\textbf{k})$. With the choice of orbital dispersions, we have $\xi^{-}_{\textbf{k}} = 2(t_{1}-t_{2})(\cos{k_x}-\cos{k_y})$, showing that the strength of $\xi^{-}_{\textbf{k}}$ is controlled by $\delta t\equiv (t_{1}-t_{2})$.


 Using the finite-q Nambu spinor, $\Psi_{\textbf{k}\textbf{q}}^{\dagger} = (\psi_{\textbf{k}}^{\dagger}, \psi_{-\textbf{k}+\textbf{q}}^{T})$, the MF Hamiltonian is given by,
\begin{equation}
    H_{\text{MF},q} = \sum_{\textbf{k}}\Psi_{\textbf{k}\textbf{q}}^{\dagger}\begin{pmatrix}
   H(\textbf{k}) & \Delta_{\textbf{q}}(\textbf{k}) \\[6pt]
   \Delta_{\textbf{q}}^{\dagger}(\textbf{k}) & -H^{*}(-\textbf{k}+\textbf{q}) \\
 \end{pmatrix}\Psi_{\textbf{k}\textbf{q}}.
 \label{hmf}
\end{equation}
Note that we assume only a single center-of-mass momentum wave vector, $\textbf{q}$, for simplicity. The OSST pairing has the form shown above, $\Delta_{\textbf{q}}(\textbf{k}) = i\tau_{2}\boldsymbol{\Delta}_{\text{OSST}}(\textbf{q})\cdot\boldsymbol{\sigma}(i\sigma_{2})$ where 
$\boldsymbol{\Delta}_{\text{OSST}}(\textbf{q})=\frac{-V}{2}\boldsymbol{d}_{a/b}(\textbf{q})$ is given in terms of the $d$ vector associated with the two orbitals $a,b$, defined as,
\begin{equation}
    \boldsymbol{d}_{a/b}(\textbf{q}) = \frac{1}{4N}\sum_{\textbf{k}\sigma\sigma\prime}[i\sigma_{2}\boldsymbol{\sigma}]_{\sigma\sigma\prime}\langle{c^{a}_{-\textbf{k}+\textbf{q}\sigma}c^{b}_{\textbf{k}\sigma\prime} - c^{b}_{-\textbf{k}+\textbf{q}\sigma}c^{a}_{\textbf{k}\sigma\prime}}\rangle.
\label{order_param}
\end{equation}
Here, an attractive interaction $V$ can be obtained from the Hund's coupling  within the MF theory \cite{klejnberg1999hund, SpalekPRB2001, HanPRB2004, Dai2008PRL, Puetter2012EPL, vafek2017hund}, as mentioned previously.

We show how the finite-q pairing state arises by self-consistently solving for the three components of the order parameter defined by Eq.~\ref{order_param} numerically, at zero temperature. 
The energy units are defined by $2t_{1}=1$ and we set $\mu=-0.4$ and $t_{ab}=0.015$. While a nonzero value for $t_{ab}$ is important to split apart the remaining orbital degeneracy in the regions of $\textbf{k}$-space where $\xi^{-}_{\textbf{k}}$ vanishes, the precise value is not important for our results. Here, we investigate the effect of a $\textbf{k}$-dependent splitting of the bands via the hopping anisotropy, $\delta t$, i.e., we tune the strength of $\xi^{-}_{\textbf{k}}$. Similar results can be obtained by adding a rigid shift of the orbitals instead, as discussed previously.

\begin{figure}[t!]
\includegraphics[width=86.5mm]{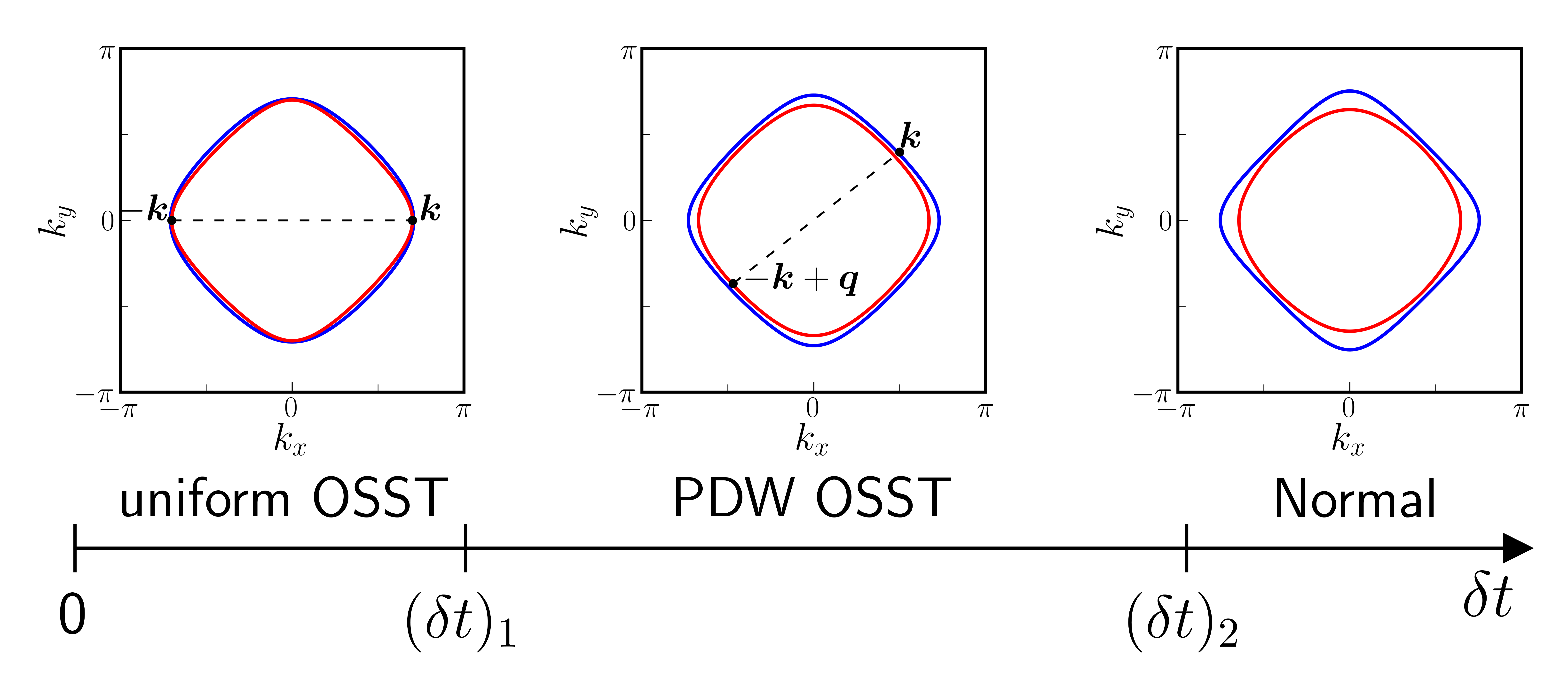}%
\caption{\label{fig1}Phase diagram for OSST pairing as a function of the orbital polarization, $\delta t$, within the two-orbital model described in the main text. Here, the orbital polarization arising from the $(d_{yz},d_{xz})$-like electronic dispersions generates a $\textbf{k}$-dependent splitting which gives rise to a transition between the uniform and finite-q OSST pairing. The first transition occurs at $(\delta t)_{1} \approx 0.016$ and the transition to the normal state at $(\delta t)_{2} \approx 0.046$. The representative Fermi surface (FS) in each phase is also shown, where the red and blue curves represent the two spin-degenerate electronic bands.}
\label{phase_diagram_1}
\end{figure}

The ordering wave vector is fixed to be of the form, $\textbf{q}=q\hat{x}$, as an examination of alternative wave vectors, including $\textbf{q}=q(\hat{x}+\hat{y})$, indicates this corresponds to the lowest free energy MF solution (see Appendix Fig.~\ref{Fig_A3} for more details). The interaction is fixed at $V=0.8$, and for each value of $\delta t$, the self-consistent solution and free energy are obtained as a function of $q$ to find the ground state solution. The resulting phase diagram as a function of $\delta t$ is shown in Fig.~\ref{phase_diagram_1}, with $(\delta t)_{1} \approx 0.016$ and $(\delta t)_{2} \approx 0.046$. For $\delta t < (\delta t)_{1}$, the uniform, i.e. $q=0$, OSST pairing is favoured due to the significant regions of orbital degeneracy in $\textbf{k}$-space. This uniform state gives way to the finite-q pairing, $q\neq 0$, for $(\delta t)_{1} <  \delta t < (\delta t)_{2}$, before the normal state is reached for $\delta t > (\delta t)_{2}$. The three components of the $d$ vector given by Eq.~\ref{order_param} are nonzero and degenerate in both phases. The order parameters as a function of $\delta t$ are shown in the Appendix (see Fig.~\ref{Fig_A2}).

Identifying solid-state candidate materials with almost degenerate Fermi surfaces (FSs) originating from two different orbitals is nontrivial due to the different dispersions of these orbitals. However, such a situation can be engineered using different degrees of freedom. For example, bilayer systems made of a single layer with one orbital exhibit almost degenerate FSs when the bilayer coupling is very weak. The resulting pairing is an interlayer singlet spin triplet, stabilized by the same form of interaction, but representing a ferromagnetic interaction between the two layers. In this case, tuning the polarization or hybridization corresponds to adjusting a layer-dependent potential or bilayer coupling, for example, via an external bias or pressure, respectively. Additionally, other systems with valley- (or sublattice-index) degrees of freedom may also be relevant, resulting in intervalley (intersublattice)-singlet spin-triplet pairing. Although we do not have specific candidate materials, our theoretical study will motivate future investigations into OSST pairing in the absence of SOC. When SOC is included, the OSST pairing does not require nearly degenerate FSs \cite{Puetter2012EPL, vafek2017hund}. Below we will include the SOC for more realistic cases and examine the resulting phases.

\subsection*{Effect of SOC}

The uniform OSST pairing is favoured when the orbital degeneracy in $\textbf{k}$-space is large, and gives way to the finite-q pairing as the degeneracy is reduced. However, in the presence of the SOC, the pairing in the OSST channel is boosted as it does not require nearly degenerate FSs. Additionally, due to SOC, intraorbital and interorbital spin-singlet pairings with the same symmetry may be induced \cite{Puetter2012EPL, vafek2017hund}. Here, we have considered a model on the square lattice with D$_{4h}$ point group, and the OSST pairing is in the $A_{1g}$ channel, which indeed contains intraorbital spin-singlet pairings \cite{Ramires2019PRB, kaba_PRB2019, Suh2019}. As the OSST pairing is the only attractive one, the intraorbital singlets are induced through the SOC, and are usually an order of magnitude smaller. Regardless of this, the introduction of SOC allows for an intraband pseudospin-singlet component of pairing from the OSST order parameter itself, which can stabilize the uniform pairing even for large orbital polarization \cite{Puetter2012EPL, vafek2017hund, Cheung2019PRB}. In light of this, one expects that the SOC and orbital polarization have opposite effects on the finite-q pairing. To demonstrate this, we now include the effects of SOC by modifying $H(\textbf{k})$ in Eq.~\ref{kinetic} to, $H(\textbf{k}) -\lambda \tau_{2}\sigma_{3}$. The SOC takes the form corresponding to the $z$ component of the atomic SOC, $2\lambda L_{z}S_{z}$.

\begin{figure}[t!]
\includegraphics[width=86.5mm]{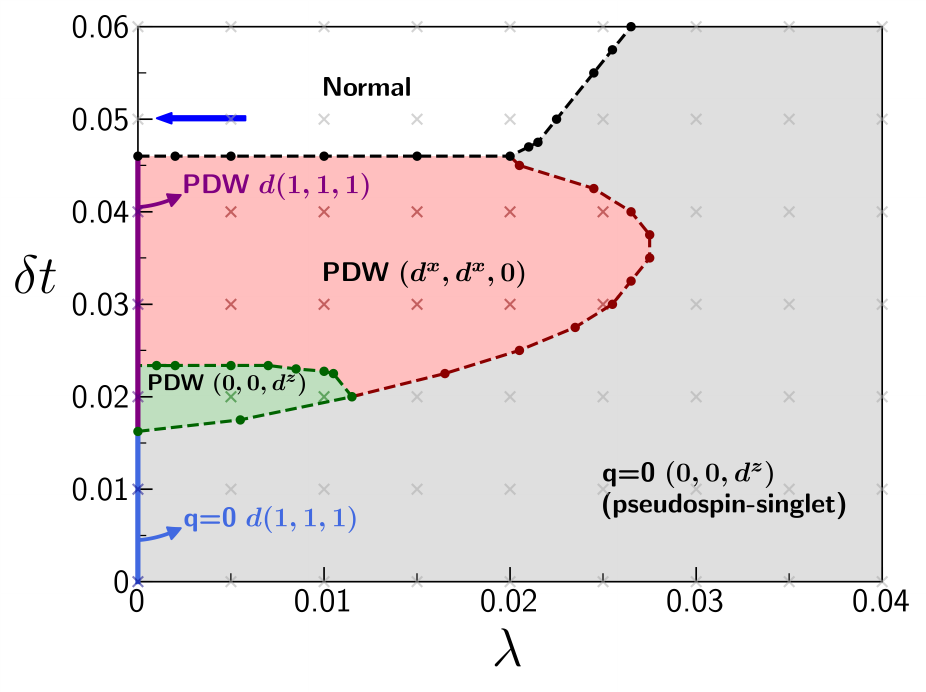}%
\caption{\label{fig2}Phase diagram for OSST pairing in the two-orbital model as a function of the orbital polarization, $\delta t$, and SOC, $\lambda$. The three components of the $d$ vector are degenerate for $\lambda=0$. For $\lambda \neq 0$, the uniform pairing (grey) corresponds to $\boldsymbol{d}_{a/b}(\textbf{q}) = (0, 0, d^{z}_{a/b}(0))$ and is generally characterized by a dominant intraband pseudospin-singlet pairing component. The smaller finite-q pairing region (green) corresponds to $(0, 0, d^{z}_{a/b}(\textbf{q}))$, while the larger region (red) corresponds to  $(d^{x}_{a/b}(\textbf{q}), d^{x}_{a/b}(\textbf{q}), 0)$. Both regions correspond to dominant interband pairing at the Fermi energy, however the former region contains a significant intraband component of pairing. This is shown in the corresponding figure in the Appendix (see Fig.~\ref{Fig_A1}). The points where calculations were done are indicated by markers, while the denser points closer to boundaries are omitted for clarity. The blue arrow represents the value of $\delta t$ used for the phase diagram shown in Fig.~\ref{phase_diagram_3}.}
\label{phase_diagram_2}
\end{figure}

We tune the strength of the SOC, $\lambda$, in addition to the orbital polarization, $\delta t$, with the resulting phase diagram shown in Fig.~\ref{phase_diagram_2}. The points where calculations were done are shown, while the denser points close to the boundaries used to determine the precise boundary locations are omitted for clarity. The $y$-axis corresponds to the phase diagram shown in Fig.~\ref{phase_diagram_1}, where the pairing is a true spin triplet and all components of the $d$ vector are degenerate, while the pairing regions for nonzero SOC are now split into those where either the $d^{z}_{a/b}$ or $\{d^{x}_{a/b}, d^{y}_{a/b}\}$ order parameters in the orbital basis are nonzero. The uniform pairing state at nonzero SOC corresponds to only the $z$-component of the $d$ vector being nonzero due to the $\sigma_{3}$ dependence of the SOC, which generates the intraband pairing component. This is labelled `$\textbf{q}=0$ pseudospin-singlet' in Fig.~\ref{phase_diagram_2} since, for large orbital polarization, it is the intraband pseudospin-singlet component which is not only dominant, but also supports the uniform pairing over the finite-q pairing. 
The finite-q pairing appears in a dome-like shape due to competing effects of $\delta t$ and SOC.

There are two separate finite-q pairing regions, with the $d$ vector rotating from the out-of-plane, $(z)$, to in-plane, $(x$-$y)$, direction when the orbital polarization is large enough. This can be explained by the $d^{x}_{a/b}(\textbf{q}),d^{y}_{a/b}(\textbf{q})$ order parameters projecting almost completely to interband  triplet pairing. In contrast, the $d^{z}_{a/b}(\textbf{q})$ order parameter contributes to both intraband singlet and interband singlet and triplet pairing. As a result, for directions parallel to $\textbf{q}$, the $(0,0,d^{z}_{a/b}(\textbf{q}))$ phase gives rise to both interband gaps at the Fermi energy, and intraband gaps close by, in contrast to the purely interband gaps in the $(d^{x}_{a/b}(\textbf{q}),d^{x}_{a/b}(\textbf{q}),0)$ phase (see Appendix A and B for more details about the MF pairing states including intraband versus interband pairing and the quasiparticle dispersion).

The finite-q  OSST pairing will therefore be favoured when the SOC is either zero, or intermediate in strength, depending on the value of the orbital polarization. As the SOC increases, the OSST pairing is no longer well-defined as the orbitals and spins are strongly coupled, and the pairing is described by an (intraband) pseudospin singlet, even though the microscopic interaction leading to the attractive interaction is defined on the original orbital and spin basis, such as the Hund's coupling. This type of spin-triplet pairing was referred to as a shadowed triplet \cite{clepkens2021,clepkens2021higher}.

The presence of the PDW phase requires a certain interaction strength, $V$, due to the imperfect nesting of the bands, a feature common to other models for PDW phases \cite{jiang_2023, Jian_PRL2015, chen_SCP2023, ticea_2024, Wu_PRL2023}. A smaller value will limit the uniform state to a smaller region of orbital polarization. This also depends on the orbital hybridization, $t_{ab}$, which is detrimental to the uniform state as mentioned previously, and the form of orbital polarization, e.g., a $\textbf{k}$-dependent versus rigid shift of the orbitals. Here, we have fixed an interaction strength to show a visible transition between the uniform and PDW state. The phase boundaries may shift depending on the details discussed above, but the qualitative description of the phase diagram remains unchanged.

\section{\label{Four} Magnetic field induced phases}

Since the SOC leads to a mixture of singlet and triplet pairing, the question of how these pairing states respond to the magnetic field naturally arises. To investigate this, we now include the effects of an in-plane Zeeman field by modifying $H(\textbf{k})$ in Eq.~\ref{kinetic} to, $H(\textbf{k}) -\lambda \tau_{2}\sigma_{3} -h_{x}\tau_{0}\sigma_{1}$, and treat $h_{x}$ as a tuning parameter. The in-plane direction for the field is chosen, as we focus on the Zeeman coupling and neglect the coupling of the orbital angular momentum to the field. 

Starting from zero SOC, and fields that are small relative to the band splitting, the field merely suppresses the $x$-component of the triplet order parameter. When the field is of the order of the band splitting, a re-entrant or field-induced pairing state with only the $x$ component of the $d$ vector may appear. Therefore, we consider a value of the orbital polarization such that there is no superconductivity, corresponding to $\delta t = 0.05$, as shown by the blue arrow in Fig.~\ref{phase_diagram_2}, and study the effects of the field and SOC together. 

\begin{figure}[t!]
\includegraphics[width=86.5mm]{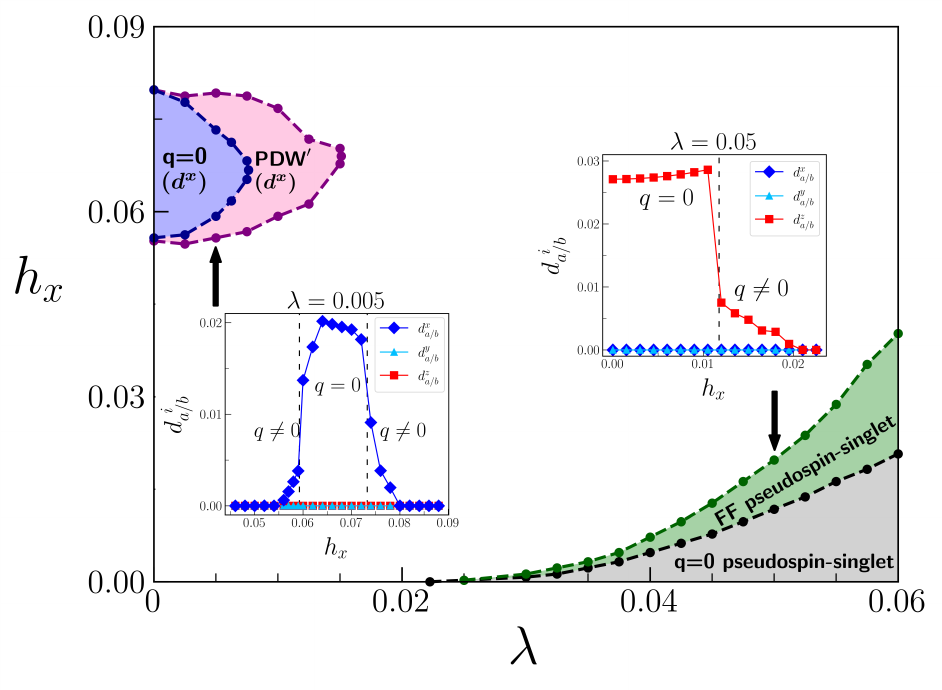}%
\caption{\label{fig3}Phase diagram as a function of magnetic field, $h_{x}$, and SOC, $\lambda$. For $\lambda \gtrsim 0.022$, the uniform state (grey) appears, and is destabilized by the field in favour of a finite-q state with a q-vector connecting the spin-split bands (green). The order parameters as a function of $h_{x}$ at $\lambda=0.05$ are shown in the right inset. There is also an orders of magnitude smaller $d^{y}_{a/b}$ component with a relative phase of $\frac{\pi}{2}$ induced by the in-plane field which is omitted for clarity. At large fields, there is a field-induced uniform pairing (blue) due to the field compensating for the orbital splitting and bringing oppositely spin-split bands together. The uniform pairing is surrounded by a finite-$q$ pairing (pink), mirroring the behaviour at low field. The order parameters are shown for $\lambda=0.005$ on the left, and the FS for $(\lambda, h_{x})=(0, 0.067)$) is shown in Fig.~\ref{Fig_C1}, along with the interband pairing amplitude.}
\label{phase_diagram_3}
\end{figure}

The resulting phase diagram is shown in Fig.~\ref{phase_diagram_3}. The magnetic field is increased from zero to a value on the same order as the band splitting on the $y$-axis, and the SOC is increased from zero along the $x$-axis. As shown in Fig.~\ref{phase_diagram_3}, there are four distinct pairing states. For values of the SOC larger than $\lambda \gtrsim 0.022$, a uniform pairing state with order parameter, $(0, 0, d^{z}_{a/b}(0))$, arises due to the SOC generating an intraband pseudospin-singlet pairing component. This pairing state behaves similarly to a true spin singlet and is therefore suppressed by the magnetic field, giving way to a finite-q pairing state with a small wave vector, $\textbf{q}=q\hat{x}$, matching the field splitting. This phase resembles a field-driven FF phase, but originates from OSST pairing. It is distinct from the zero-field PDW phase, which corresponds to interband pairing between two Kramer's degenerate bands, rather than between two spin-split bands. The order parameters are plotted as a function of the magnetic field in the inset on the right for $\lambda=0.05$. A jump corresponding to the transition from $q=0$ to $q \neq 0$ can be seen. In addition to the primary out-of-plane spin-triplet order parameter, an orders of magnitude smaller $d^{y}_{a/b}$ component with a relative $\frac{\pi}{2}$ phase is induced by the in-plane field, which we omit for clarity.

When the magnetic field approaches the band splitting caused by the orbital polarization and hybridization, the two bands which are oppositely spin-polarized approach each other in $\textbf{k}$ space. This is shown by the representative FS for $(\lambda, h_{x}) = (0, 0.067)$ in the Appendix (see Fig.~\ref{Fig_C1}). When the SOC is small, the field reduces the band splitting enough to drive a transition to the uniform pairing with order parameter, $(d^x_{a/b}(0),0,0)$, i.e., pairing between up- and down-spin along the $x$ direction. As the SOC increases, due to the non-commuting nature of the Zeeman field proportional to $\sigma_{1}$ and the SOC proportional to $\sigma_{3}$, the bands do not approach each other sufficiently close enough for the uniform pairing to emerge. Instead, a window of finite-q pairing with order parameter, $(d^x_{a/b}(q),0,0)$, is obtained, where $q$ is given by the mismatch between the two oppositely spin-polarized bands. This phase is labelled by $\text{PDW}^{\prime}$ in Fig.~\ref{phase_diagram_3} since it occurs in the presence of time-reversal symmetry breaking but is distinct from an FF phase. As the SOC increases further, this pairing phase is also suppressed. The order parameters as a function of the field are shown in the inset on the left in Fig.~\ref{phase_diagram_3} for $\lambda=0.005$. The order parameters correspond to a cut through the region with the uniform pairing surrounded by two finite-q pairing states.

Since the SOC works together with the Hund's coupling to produce a weak-coupling instability \cite{Puetter2012EPL, vafek2017hund}, the existence of the uniform phase is robust to changes in the interaction strength. However the region of the field-induced phases is expected to shrink/expand for smaller/larger interaction strengths, which also depends on the orbital hybridization and polarization, as discussed above.

A similar field-induced uniform interband pairing has been found recently in Refs.~\cite{salamone_PRB2023,chakraborty_2023}, starting from purely spin-singlet pairing. In Ref.~\cite{chakraborty_2023}, a single-band system was considered, with the field-induced phase requiring the presence of altermagnetism. Ref.~\cite{salamone_PRB2023} considered an effective two-band conventional spin-singlet SC with both intra and interband pairing. Here, we have shown that OSST pairing exhibits a similar uniform field-induced phase, and that such a phase can survive inclusion of the SOC up to a critical value which will depend on the exact microscopic model. However, in addition to the uniform field-induced phase, we also find a finite-q pairing state which is induced by a large magnetic field. The transitions between the finite-q and uniform phases at large field mirror the FF-like pairing phase transition for larger values of SOC. For the former, the pairing wave vector connects oppositely spin-polarized bands that originate from two distinct zero-field Kramer's degenerate bands. Concretely, the pairing in the band basis, defined by fermionic operators, $f^\dagger_{i,\textbf{k},s}$, in bands $i=1,2$ with pseudospin $s=+/-$,  corresponds to interband pairing, $\langle{f^\dagger_{1,\textbf{k},-}f^\dagger_{2,-\textbf{k}+\textbf{q},+}}\rangle \neq 0$. With negligible SOC, the spins $+$/$-$ are defined by the direction of the field. In the FF-like pairing state, the dominant pairing component in the band basis for small fields corresponds to intraband pairing, i.e.,  $\langle{f^\dagger_{i,\textbf{k},-}f^\dagger_{i,-\textbf{k}+\textbf{q},+}}\rangle \neq 0$, due to the SOC.

\section{\label{Five}finite-q pairing in  \protect\Sr}

The pairing phases can be investigated in other multiorbital systems beyond the two-orbital case. For instance, with three $t_{2g}$ orbitals, there are now three interorbital $d$ vectors with the same effective interaction originating from Hund's coupling, which we investigate next. The complexity can be further enhanced by the mixing between the three orbitals, allowing for nontrivial momentum-dependent pairings \cite{Suh2019, clepkens2021, clepkens2021higher}. Since our aim is to address finite-q pairing, we restrict ourselves to the $A_{1g}$ state stabilized by atomic SOC \cite{Puetter2012EPL, vafek2017hund}. While details such as the direction of the $q$-vector may depend on the precise OSST pairing symmetry, the existence of the finite-q state itself is not contingent upon the pairing symmetry.

We consider three $t_{2g}$ orbitals on a square lattice, along with sizeable atomic SOC and Hund's coupling. Depending on the tight-binding parameters used, such a model can describe Sr$_{2}$RuO$_{4}$, and other materials such as the iron-pnictide SCs \cite{daghofer_PRB2010}.  Here, we consider the application to Sr$_{2}$RuO$_{4}$ and show that OSST pairing can give rise to a finite-q pairing state. While our aim is not to perform a detailed study of Sr$_{2}$RuO$_{4}$, we note that OSST pairing is one of the candidate order parameters for Sr$_{2}$RuO$_{4}$, making this a relevant application \cite{Suh2019, clepkens2021higher, maeno_2024}.

We consider an effective interaction Hamiltonian containing an attractive channel for OSST pairing originating from the on-site Hubbard-Kanamori interaction terms \cite{klejnberg1999hund,SpalekPRB2001,HanPRB2004,Dai2008PRL,Puetter2012EPL, vafek2017hund},
\begin{equation}
    H_{\text{eff}} = -2NV\sum_{\alpha< \beta}\hat{\boldsymbol{d}}_{\alpha/\beta}^{\dagger}(\textbf{q})\cdot\hat{\boldsymbol{d}}_{\alpha/\beta}(\textbf{q}),
\end{equation}
where $N$ is the number of sites and $\alpha< \beta$ represents a sum over the three unique pairs of orbitals. The interaction strength, $V = J_{H} - U^{\prime}$, is given in terms of the renormalized low-energy interorbital Hubbard ($U'$) and Hund's interaction ($J_H$) parameters \cite{Puetter2012EPL, vafek2017hund}. While the condition for an attractive interaction in MF theory is unrealistic, studies going beyond MF theory have found OSST pairing with realistic bare interaction parameters \cite{Hoshino2015PRL, Hoshino2016PRB,Gingras2019PRL}. Indeed, the local effective interaction between same-spin electrons becomes attractive within second-order perturbation theory when the local magnetic susceptibility is large \cite{Inaba_PRL2012,Hoshino2015PRL}. We calculate the nine order parameters, $\boldsymbol{d}_{\alpha/\beta}$, self-consistently and consider only a single wave vector, $\textbf{q}$, in the MF theory to simplify the calculation, as the qualitative physics is not expected to change from this choice.

We use the normal-state Hamiltonian described in Ref.~\cite{lindquist2019distinct}, and described in the Appendix, which includes a sizeable value of SOC, appropriate for Sr$_{2}$RuO$_{4}$ \cite{Tamai2019PRX}. We express energies in units of $2t_{3}=1$, where $t_{3}$ is the nearest-neighbour hopping of the $d_{xy}$ orbital. In these units, the SOC strength is $\lambda=0.085$. Note that the difference with our previous model is that now the orbital polarization between $(d_{yz}$,~$d_{xz})$ orbitals is much larger, and we have the complete atomic SOC, $2\lambda\textbf{L}\cdot\textbf{S}$, acting in the space of the three $t_{2g}$ orbitals. An external in-plane Zeeman field is included, $H_{\text{Zeeman}} = -gh_{x}\sum_{i}S^{x}_{i}$, with $g=2$ assumed for simplicity. Fixing the interaction strength at $V=0.6$, we self-consistently solve for the order parameters, $\boldsymbol{d}_{\alpha/\beta}(\textbf{q})$, at zero temperature as a function of $\textbf{q}$. We choose an interaction strength such that the uniform state is visible up to a reasonably sized field value for resolving the $q$-vector in the finite-q phase. As before, wave vectors of the form, $\textbf{q} = q\hat{x}$, are found to be more stable than $\textbf{q} = q(\hat{x}+\hat{y})$.

\begin{figure}[t!]
\includegraphics[width=86.5mm]{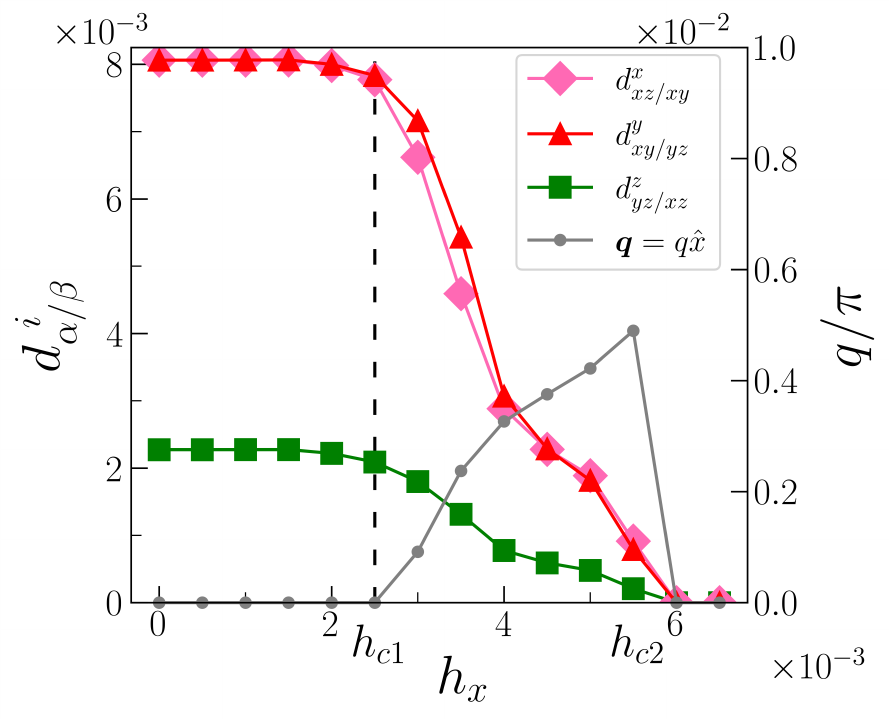}%
\caption{\label{fig4}MF order parameters, $\boldsymbol{d}_{\alpha/\beta}$, and ordering wave vector, $\textbf{q}=q\hat{x}$, for the three-orbital model \cite{lindquist2019distinct} as a function of in-plane magnetic field, $h_{x}$. Of the nine order parameters, only the three nonzero ones are shown for clarity. The in-plane magnetic field along the $x$-direction also induces the $d^{z}_{xy/yz}$ and $d^{y}_{yz/xz}$ order parameters to become nonzero with a relative phase of $\frac{\pi}{2}$, however they are orders of magnitude smaller and thus omitted. The two critical fields denoting the transition to the finite-q and normal state are, $h_{c1}\approx 0.003$ and $h_{c2}\approx0.006$, respectively.}
\label{MF_SRO}
\end{figure}

The resulting order parameters, along with the associated wave vectors, are shown in Fig.~\ref{MF_SRO}. Starting from zero, the in-plane field, $h_{x}$, is increased along the $x$-axis. At zero field, the atomic SOC favours a uniform pairing state with the three primary nonzero order parameters, $(d^{x}_{xz/xy}(0), d^{y}_{xy/yz}(0), d^{z}_{yz/xz}(0))$. The $d^{z}_{yz/xz}$ pairing corresponds to the SOC-stabilized order parameter in our two-orbital model shown previously, while the other two involving the $d_{xy}$ orbital carry the same $A_{1g}$ symmetry. The in-plane magnetic field also induces the $d^{z}_{xy/yz}$ and $d^{y}_{yz/xz}$ order parameters to become finite with a relative phase of $\frac{\pi}{2}$, however they are orders of magnitudes smaller. In the band basis where orbitals and spins are mixed, the intraband pseudospin-singlet is dominant, as discussed above. This is apparent from the suppression of all three components of the $d$ vector as the field is increased from zero. At $h_{x}\approx h_{c1}$, the pair-breaking effect of the field is large enough such that a finite-q pairing develops, and there is a transition to an FF phase. At this point, the $x$ component of the $d$ vector is noticeably suppressed compared to the $y$ component, as expected for the field along the $x$ direction for a triplet order parameter. 
The wave vector increases with the field until the normal state is reached at $h_{x} \approx h_{c2}$. 

The results shown here with three orbitals are similar to the low-field pairing phases shown in Fig.~\ref{phase_diagram_3} for the simpler two-orbital model, although the presence of the third orbital leads to two additional OSST order parameters. The presence of finite-q superconductivity in the case of moderately large SOC is therefore robust to additional complexity introduced when describing a $t_{2g}$-orbital material such as Sr$_{2}$RuO$_{4}$. As shown by Fig.~\ref{phase_diagram_3}, the high-field phases are expected to be relevant for a material with sizable band degeneracy and small SOC. The application of these parts of the parameter space, as well as those found in Fig.~\ref{phase_diagram_2} for other multiorbital SCs such as the Fe-based SCs are left for future study.

\vspace{-3mm}
\section{\label{Six} Summary and Discussion}

In summary, we have shown how finite-q pairing, referred to as PDW or FF phases depending on the origin of the finite center-of-mass momentum, can be generated from uniform OSST pairing. With zero external field, the PDW state arises due to the pair-breaking effects of the orbital polarization and/or hybridization on the uniform state. Using a simple model of two orbitals on a square lattice, we demonstrate the appearance of the PDW upon increasing the orbital polarization. When the SOC is introduced, the PDW region splits into two separate phases with different $d$-vector directions due to the competition between interband pairing characterized by the $d$ vector in the $x-y$ plane, versus a combination of intraband and interband pairing characterized by the $d$ vector along the z-axis. An external in-plane magnetic field induces the uniform OSST pairing at high fields for small SOC, which becomes a finite-q pairing for larger SOC. Upon increasing SOC, the uniform pseudospin-singlet pairing is found at low fields, which becomes the FF phase by increasing the field. We have applied this picture to the three-orbital SC, Sr$_{2}$RuO$_{4}$, which has exhibited signs of an FFLO phase in recent nuclear magnetic resonance experiments \cite{Kinjo_Science2022}. Within a three-orbital model for the $t_{2g}$ orbitals and SOC, the uniform OSST pairing evolves into an FF phase with an in-plane Zeeman field. This phase is characterized by three distinct spin-triplet components with the associated $d$ vectors.
Due to the strong SOC, the dominant pairing can be described as an intraband pseudospin-singlet. 

Our analysis focuses solely on the finite-q $A_{1g}$ pairing with a single wave vector. However, in materials like Sr$_{2}$RuO$_{4}$, pairing symmetries with higher angular momentum, such as d-wave ($B_{1g}$ or $B_{2g}$), have been proposed. We expect that the theoretical framework we are employing for finite q-pairing in the presence of SOC and in-plane magnetic fields can also be extended to accommodate these higher angular momentum pairing states. This parallels the situation observed in the FFLO state for a d-wave spin singlet under the Zeeman field \cite{shimahara_JPSJ1997,Yang_PRB1998}, which bears similarities to the s-wave spin singlet case.

A few additional aspects concerning the limitations of the current study and topics for future investigation warrant some discussion. For a two-orbital model, we have used the component of the atomic SOC, $L_{z}S_{z}$, relevant for the $d_{xz}$ and $d_{yz}$ orbitals. In this case, the phase diagram depends on the direction of the field. We have presented the phase diagram for the field along the $x$ (or equivalently $y$)-direction.
When the field aligns with the $z$-direction, the uniform field-induced phase may extend to larger values of SOC. In this scenario, a transition between multiple uniform and finite-q phases separated by a large field interval may be possible. However, if this corresponds to the out-of-plane direction, the orbital limiting of the pairing will have to be considered. We note that in the spin-triplet SC, UTe$_{2}$, there is a rich phase diagram as a function of magnetic field, which exhibits low- and high-field pairing phases \cite{aoki_JPCM2022}, in addition to recent evidence for finite-q superconductivity \cite{gu_Nat2023, aishwarya_Nat2023}. The application of OSST pairing to this problem is left as an interesting future direction.

An additional candidate for finite-q pairing originating from an OSST order parameter is the family of Fe-based SCs. The OSST pairing state has been proposed previously to explain the superconductivity of the iron-pnictide SCs \cite{Dai2008PRL, vafek2017hund, Coleman_PRL2020}, due to the strong Hund's coupling and significant SOC \cite{borisenko_Nat2016}. Indeed, a PDW state based on OSST pairing was also proposed using a two-band model for LaFeAsO$_{1-x}$F$_{x}$ \cite{zegrodnik_JSNM2015}. However, the SOC and associated intraband pairing was neglected in Ref.~\cite{zegrodnik_JSNM2015}. We have shown that the finite-q PDW phase within our model can survive the inclusion of SOC, and leads to two different PDW states. Furthermore, we have found the presence of finite-q and uniform field-induced phases for a range of SOC values up to  $3\%$ of the largest NN hopping parameter. The SOC strength for the Fe-based  SCs may be of an appropriate size for these phases to become relevant.  However, whether an OSST PDW phase appears in the iron pnictides taking into account the SOC and intersite interactions, as well as competition with other order parameters such as a spin-density wave remains as a future study.


Finally, while we have neglected finite-frequency effects in the present MF study, it was pointed out that odd-frequency pairing correlations are ubiquitous in multiband systems \cite{AnnicaPRB2013, Komendova_PRL2017}. For example, an odd-frequency odd-interband pairing component will be induced from even-frequency even-interband pairing. In our scenario, such a pairing will therefore already be present for $\textbf{q}=0$, when both SOC and orbital hybridization are finite, and the finite-q pairing is expected to lead to additional odd-frequency correlations. Such odd-frequency components may have important experimental consequences, for example in the superfluid weight \cite{Chakraborty_PRB2022}. More generally, the critical role of finite-frequency correlation effects on superconductivity and other instabilities in a Hund's metal have become apparent from dynamical MF theory studies \cite{Fanfarillo_PRL2020, Fanfarillo_PRB2023}. In particular, when dynamical correlations are taken into account, the Hund's coupling allows for the superconducting state to survive in the strong correlation limit, despite a small quasiparticle weight, which is not captured by a renormalized quasiparticle approximation  \cite{Fanfarillo_PRL2020}. Several dynamical MF theory studies for Sr$_{2}$RuO$_{4}$ have also found pairing states containing both even- and odd-frequency pairing \cite{Gingras2019PRL, Gingras_PRB2022}, and have shown that the finite-frequency correlations are essential for describing the normal-state properties \cite{mravlje2011PRL, Kugler_PRL2020, Blesio_PRR2024}. An intriguing future avenue of investigation is therefore the question of what role finite-frequency correlations may play in OSST finite-q states in the Hund's metal regime, which may be studied by one of the cluster extensions of dynamical MF theory \cite{Lichtenstein_PRB2000, Kotliar_PRL2001}.

\vspace{1em}

\begin{acknowledgments}
This work was supported by the Natural Sciences and Engineering Research Council of Canada (NSERC) Discovery Grant No. 2022-04601, the Center for Quantum Materials at
the University of Toronto, the Canadian Institute for
Advanced Research, and the Canada Research Chairs Program. Computations were performed on the Niagara supercomputer
at the SciNet HPC Consortium. SciNet is funded by Innovation, Science and Economic Development Canada; the Digital
Research Alliance of Canada; the Ontario Research Fund:
Research Excellence; and the University of Toronto.
\end{acknowledgments}

\section*{Appendix A: MF pairing states at zero field}

\renewcommand{\theequation}{A\arabic{equation}}
\setcounter{equation}{0}
\renewcommand{\thesubsection}{\arabic{subsection}}
\renewcommand\thefigure{A\arabic{figure}} 
\setcounter{figure}{0}
\setcounter{table}{0}
\renewcommand{\thetable}{A\arabic{table}}

\renewcommand{\theHfigure}{A\arabic{figure}}

\begin{figure}[t!]
\includegraphics[width=86.5mm]{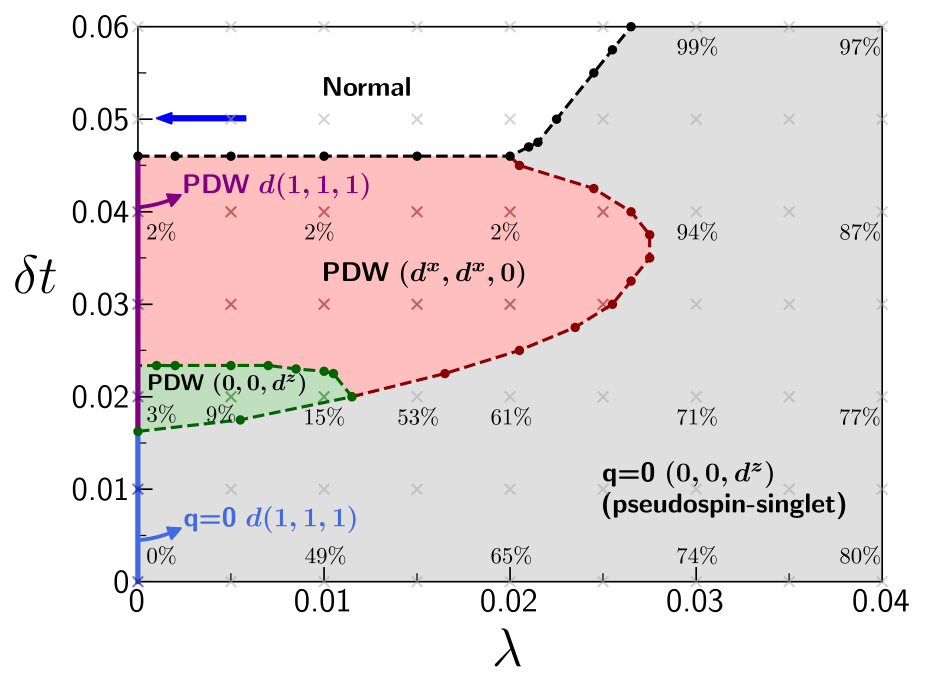}%
\caption{Phase diagram corresponding to Fig.~\ref{fig2} of the main text, with the percentage of intraband pairing added to representative points. Due to the finite-q, both PDW phases can acquire a tiny percentage of intraband pairing when the SOC is zero. Since the SOC does not induce intraband pairing for the $(d_{a/b}^{x}, d_{a/b}^{x}, 0)$ state, this remains constant as $\lambda$ increases, while it increases in the $(0,0,d_{a/b}^{z})$ state (see text for more details).}
\label{Fig_A1}
\end{figure}

The phase diagram as a function of SOC, $\lambda$, and orbital polarization, $\delta t$, is reproduced in Fig.~\ref{Fig_A1}, with the percentage of intraband pairing shown at representative points. The percentage of intraband pairing in the MF pairing solutions is defined by, $\frac{P_{\text{intra}}}{P_{\text{intra}} + P_{\text{inter}}}$, where the FS-averaged total intraband pairing amplitude is $P_{\text{intra}} =\frac{1}{2N_{F}}\sum_{iss^{\prime},\textbf{k}\in\text{FS}}|\langle{f_{i,\textbf{k},s}f_{i,-\textbf{k}+\textbf{q},s^{\prime}}}\rangle|$. The number of FS points is $N_{F}$, and cancellations due to a \textbf{k}-dependent sign are avoided by taking the absolute value at each point. Similarly the total interband pairing amplitude is, $P_{\text{inter}} =\frac{1}{2N_{F}}\sum_{i\neq j,ss^{\prime},\textbf{k}\in\text{FS}}|\langle{f_{i,\textbf{k},s}f_{j,-\textbf{k}+\textbf{q},s^{\prime}}}\rangle|$.

\begin{figure}[t!]
    \centering
\begin{tabular}{cc}
  \includegraphics[width=41mm]{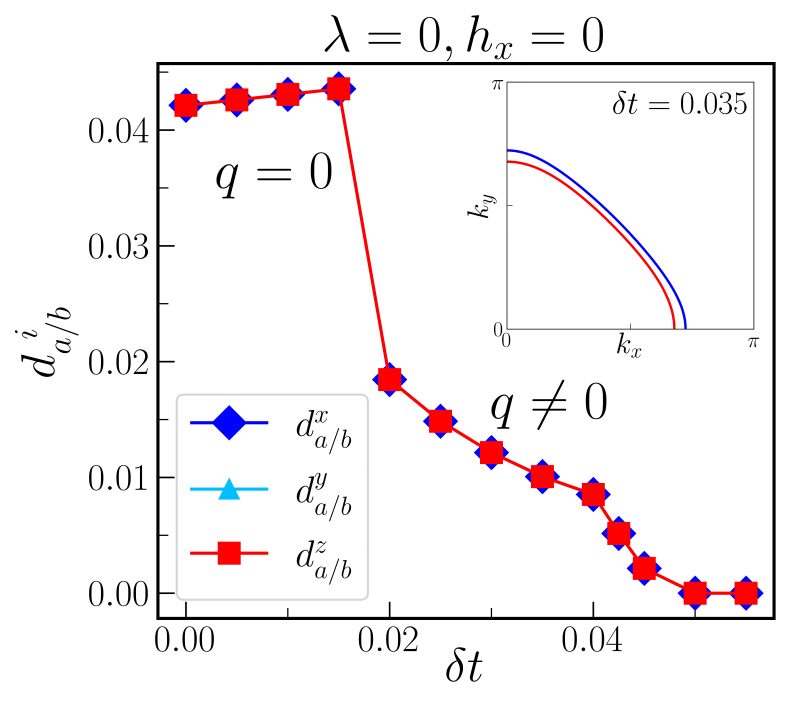} &  \includegraphics[width=41mm]{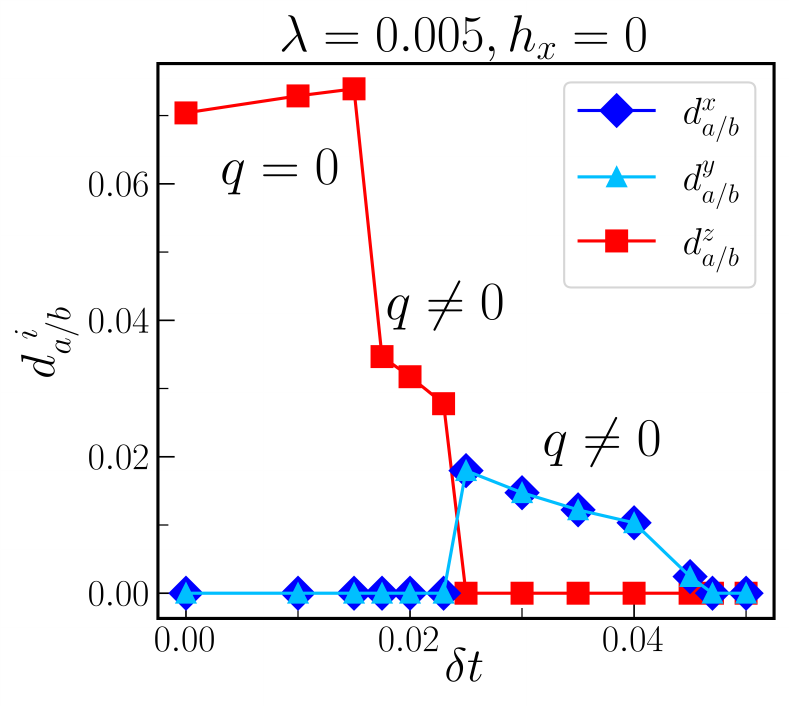}
\end{tabular}
\caption{MF order parameters corresponding to cuts at $\lambda=0$ and $\lambda=0.005$ through Fig.~\ref{Fig_A1}.} 
\label{Fig_A2}
\end{figure}

The MF order parameters corresponding to the three components of Eq.~\ref{order_param} are shown for two different cuts through the phase diagram in Fig.~\ref{Fig_A2}. The left (right) panels correspond to fixed values of $\lambda = 0 (0.005)$ respectively, and zero magnetic field, $h_{x}=0$. A representative FS for $\lambda=0$ and a value of $\delta t = 0.035$, where the $\textbf{q}\neq 0$ state is stabilized, is shown as an inset in the left panel. The ordering wave vector is of the form, $\textbf{q} = q\hat{x}$, which best connects the two bands near the X point as discussed below. A lattice size of $1600\times1600$ was used, which determines the resolution in $q$, and we have checked larger sizes to confirm our results are qualitatively unchanged.

Since the SOC is zero, the OSST pairing corresponds to purely interband spin-triplet pairing in the band basis for $q=0$. The right panel shows the two separate $q \neq 0$ phases with different $d$-vector directions that appear in Fig.~\ref{Fig_A1} for small SOC. There is a slight increase in the order parameters before the development of a finite $q$ as $\delta t$ is increased. We attribute this to an increase in the density of states at the Fermi energy due to a splitting, $\propto 4\delta t$, of the two van Hove singularities at higher energy. In the finite-q state, there is a smaller average gap, since the q-vector doesn't allow for a gap over all regions of the FS.

The difference between the in-plane ($d^{x}_{a/b}, d^{y}_{a/b}$) and out-of-plane ($d^{z}_{a/b}$) order parameters is that, for $q=0$, the in-plane one projects to purely interband spin-triplet pairing, while the out-of-plane one projects to both interband spin-triplet, interband spin-singlet, and intraband spin-singlet. When $q \neq 0$, both states may also acquire an additional small intraband pairing component due to contributions from kinetic terms of the form, $(H^{\gamma\gamma^{\prime}}(\textbf{k})-H^{\gamma\gamma^{\prime}}(-\textbf{k} + \textbf{q}))$ (for example from the orbital hybridization, $t_{\textbf{k}}=H^{10}(\textbf{k})$). This is shown by the tiny percentage ($\approx 2-3\%$) of intraband pairing in the $q\neq 0$ phase for $\lambda=0$. For small values of orbital polarization, the $d_{a/b}^{z}(\textbf{q})$ triplet order parameter allows for an intraband component of pairing to exist near the Fermi energy as SOC is increased (shown also by the quasiparticle dispersion below in Fig.~\ref{Fig_B1}). However once the splitting between bands is large enough, $\delta t \approx 0.025$, the $(d_{a/b}^{x}, d_{a/b}^{x}, 0)$ pairing state becomes more favourable due to a larger interband pairing component.

\begin{figure}[h!]
    \centering
\begin{tabular}{c}
  \includegraphics[width=70mm]{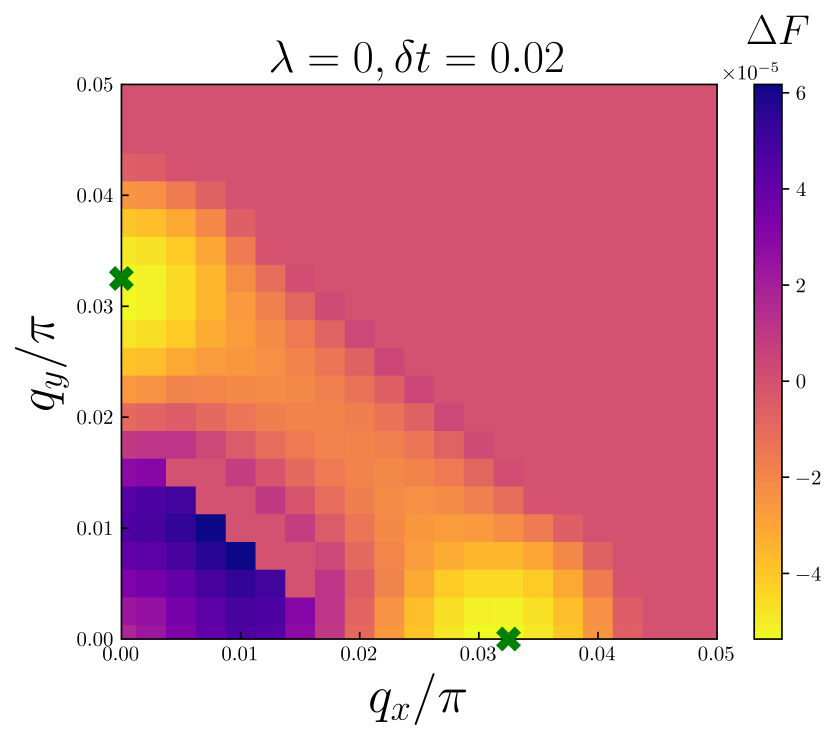} 
\end{tabular}
\caption{Free energy difference with the normal state, $\Delta F = F_{\text{MF}}-F_{\text{N}}$, for self-consistent solutions obtained over a grid of wave vectors at a representative point in the parameter space, $(\lambda=0, \delta t=0.02)$. The best MF solution is obtained for the two degenerate wave vectors of the form, $\textbf{q} = (q,0)$ or $\textbf{q} = (0,q)$, which is indicated by the green markers. For computational efficiency, a smaller system size of $800\times800$ was used for this calculation.} 
\label{Fig_A3}
\end{figure}

\renewcommand\thefigure{B\arabic{figure}} 
\setcounter{figure}{0}
\renewcommand{\theHfigure}{B\arabic{figure}}


\begin{figure*}[t!]
    \centering
\begin{tabular}{cc}
  \includegraphics[width=170mm]{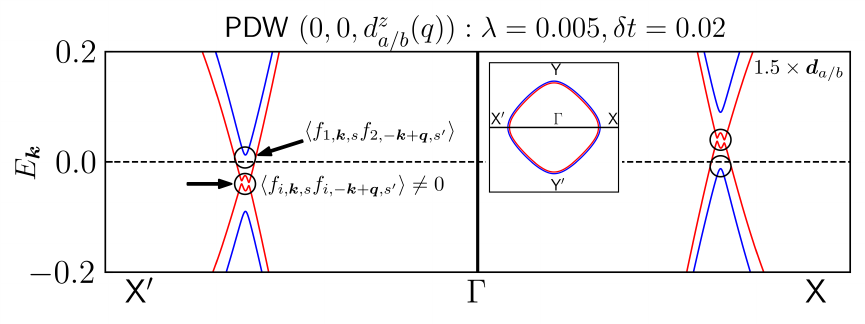} 
\end{tabular}
\caption{Quasiparticle dispersion for the MF solution, $(0, 0, d^{z}_{a/b}(\textbf{q}))$, corresponding to $\lambda=0.005$ and $\delta t = 0.02$. The $\text{X}^{\prime}-\Gamma-\text{X}$ direction is shown, depicted by the inset FS, which is parallel to the pairing wave vector, $\textbf{q}=q\hat{x}$.} 
\label{Fig_B1}
\end{figure*}

\begin{figure*}[t!]
    \centering
\begin{tabular}{cc}
  \includegraphics[width=170mm]{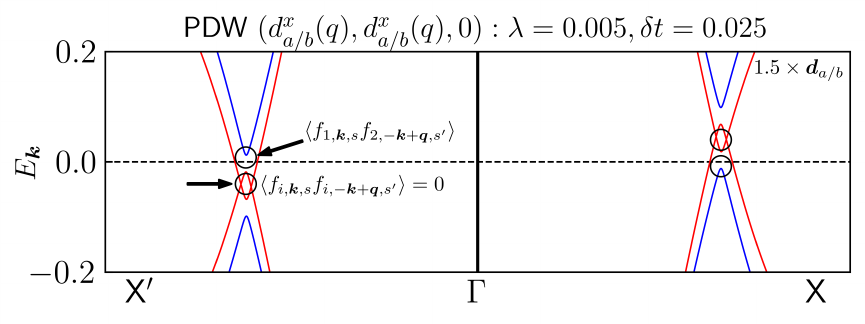}
\end{tabular}
\caption{Quasiparticle dispersion for the MF solution, $(d^{x}_{a/b}(\textbf{q}), d^{x}_{a/b}(\textbf{q}), 0)$, corresponding to $\lambda=0.005$ and $\delta t = 0.025$.} 
\label{Fig_B2}
\end{figure*}

\begin{figure*}[t!]
    \centering
\begin{tabular}{cc}
  \includegraphics[width=88.2mm]{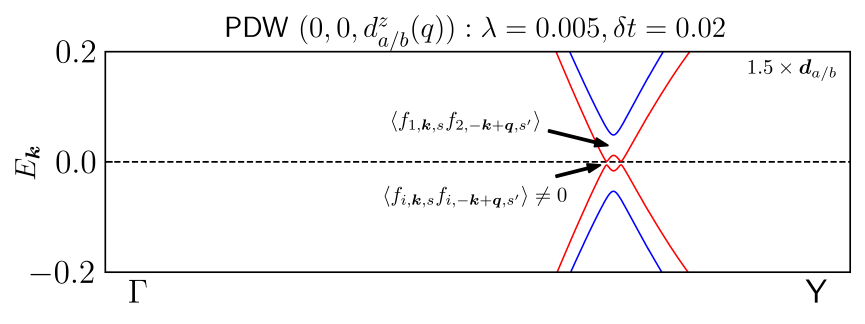} &  \includegraphics[width=88.2mm]{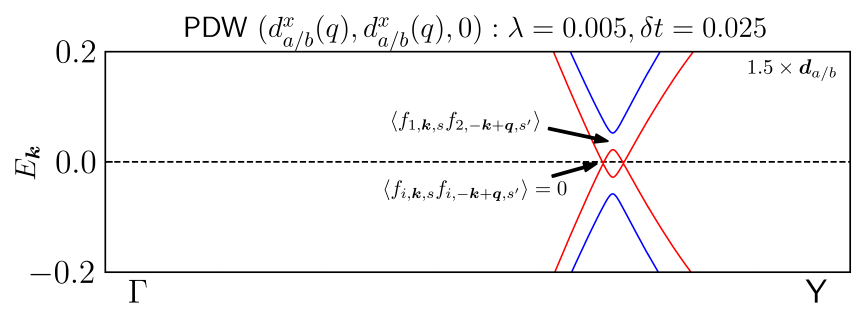}
\end{tabular}
\caption{Quasiparticle dispersion for the MF solutions corresponding to $\lambda=0.005$ and $\delta t = 0.02 (0.025)$ in the left(right) panels for the $\Gamma-\text{Y}$ direction, which is perpendicular to $\textbf{q}$. Only the positive $\textbf{k}$ direction is shown for the $y$-direction since the dispersion is symmetric.} 
\label{Fig_B3}
\end{figure*}

The free energy difference between the MF solutions and the normal state, as a function of $\textbf{q}=(q_{x},q_{y})$ for a representative point, $(\lambda=0, \delta t=0.02)$, is shown in Fig.~\ref{Fig_A3}. As discussed in the main text, the best solution is found for wave vectors of the form, $\textbf{q}=(q,0)$ or $\textbf{q}=(0,q)$, with the two being degenerate due to the $C_{4}$ symmetry. We have also checked the comparison between $(q,0)$ and $(q,q)$ wave vectors for other points in the parameter space and find the same result. We attribute this finding to the better nesting of the bands along the directions parallel to $x$ or $y$ for such a $q$-vector, which is more favourable for pairing due to a larger density of states from nearby van Hove singularities at the X/Y points of the Brillouin zone.

\section*{Appendix B: Quasiparticle dispersion}

\renewcommand{\theequation}{B\arabic{equation}}
\setcounter{equation}{0}
\setcounter{table}{0}
\renewcommand{\thetable}{B\arabic{table}}




The quasiparticle dispersion obtained by diagonalizing Eq.~\ref{hmf} of the main text for the MF solution with $\lambda=0.005$ and $\delta t=0.02$ (Fig.~\ref{Fig_B1}) as well as $\delta t=0.025$ (Fig.~\ref{Fig_B2}) are shown. The former solution corresponds to the state with out-of-plane $d$ vector, $(0, 0, d^{z}_{a/b}(\textbf{q}))$, while the latter corresponds to the in-plane $d$-vector state, $(d^{x}_{a/b}(\textbf{q}), d^{x}_{a/b}(\textbf{q}), 0)$. The dispersions are shown plotted over the direction in $\textbf{k}$-space parallel to the wave vector, $\textbf{q}=q\hat{x}$, i.e., the $\text{X}^{\prime}-\Gamma-\text{X}$ direction, as depicted in the inset FS of Fig.~\ref{Fig_B1}. The finite $\textbf{q}$ gives rise to an overlap between a particle $f_{i}$-band and the corresponding shifted hole $f_{j}$-band ($i\neq j)$ at the Fermi energy for positive and negative $\textbf{k}$. This opens up a gap originating from the interband pairing, $\langle{f_{i,\textbf{k},s}f_{j,-\textbf{k}+\textbf{q},s^{\prime}}}\rangle$, as shown by the circles at the Fermi energy in Fig.~\ref{Fig_B1} and Fig.~\ref{Fig_B2}. However, due to the single-$\textbf{q}$, an unpaired set of bands arise for both positive and negative $\textbf{k}$, leading to the depaired regions \cite{Fulde1964}. The intraband crossings are shifted away from the Fermi energy, and are gapped (gapless) for $\delta t = 0.02 (0.025)$ as shown by the circles in both figures. This is due to the intraband component of pairing, $\langle{f_{i,\textbf{k},s}f_{i,-\textbf{k}+\textbf{q},s^{\prime}}}\rangle$, which is significant only for the out-of-plane $d$-vector pairing state. 

The dispersion in Fig.~\ref{Fig_B3} shows the direction perpendicular to $\textbf{q}$, i.e., the $\Gamma-\text{Y}$ direction for both pairing states. Only the positive $\textbf{k}$ direction is shown for clarity since the dispersion is symmetric. Since this direction is perpendicular to $\textbf{q}$, the two distinct bands near the $\text{Y}$ point are not connected to each other after being shifted, but rather mostly overlap with themselves. Therefore, the interband gaps occur away from the Fermi energy, similar to the $\textbf{q}=0$ case. The intraband crossings are shifted slightly away from the Fermi energy and are gapped (gapless) for $\delta t=0.02 (0.025)$.

\onecolumngrid

\section*{Appendix C: Field-induced pairing}

\renewcommand{\theequation}{C\arabic{equation}}
\setcounter{equation}{0}
\renewcommand\thefigure{C\arabic{figure}} 
\setcounter{figure}{0}
\setcounter{table}{0}
\renewcommand{\thetable}{C\arabic{table}}
\renewcommand{\theHfigure}{C\arabic{figure}}

\begin{figure}[h!]
    \centering
\begin{tabular}{cc}
  \includegraphics[width=90mm]{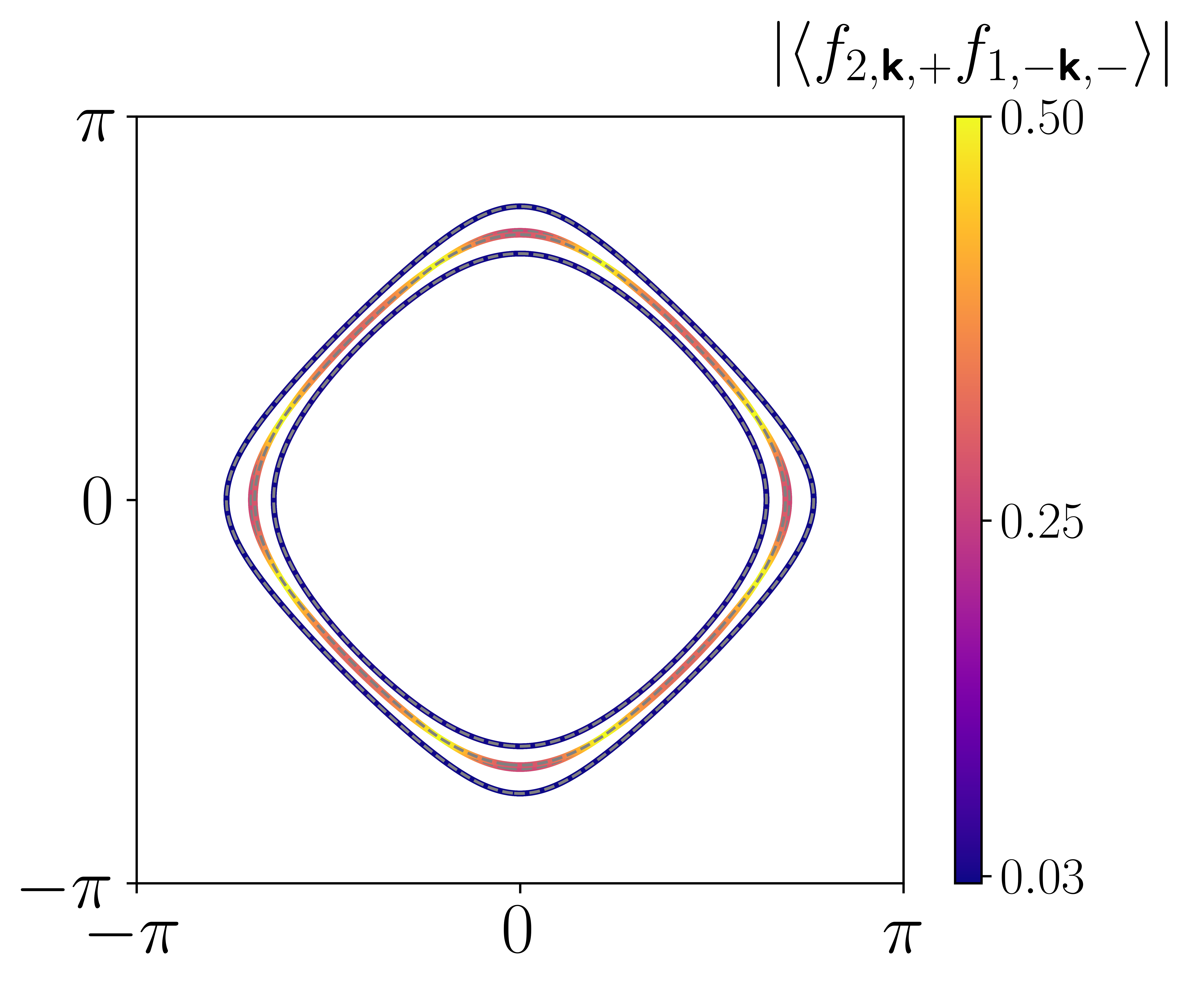} 
\end{tabular}
\caption{Interband pairing amplitude for the field-induced pairing state with order parameter, $(d^{x}_{a/b}(0), 0, 0)$, plotted over the normal-state FS shown by the grey dashed lines, for $(\lambda, \delta t, h_{x})=(0, 0.05, 0.067)$.} 
\label{Fig_C1}
\end{figure}

The field-induced pairing region shown in Fig~\ref{phase_diagram_3} of the main text arises due to the oppositely spin-polarized bands approaching each other in $\textbf{k}$-space as the magnetic field is increased. For this to occur, the pairing in the orbital basis must project to interband pairing. For zero SOC, the uniform OSST order parameters projects to purely interband pairing, making the field-induced phase robust. To show this, the interband pairing amplitude, $\langle{f_{2,\textbf{k},+}f_{1,-\textbf{k},-}}\rangle$ is shown in Fig.~\ref{Fig_C1} for a representative field value for zero SOC in the uniform field-induced region, $(\lambda, \delta t, h_{x}) = (0, 0.05, 0.067)$, with order parameter, $(d^{x}_{a/b}(0), 0, 0)$. The underlying normal-state FS is shown by the grey dashed line. The pairing is largest on the two overlapping oppositely spin-polarized bands, $f_{2,\textbf{k},+}, f_{1,\textbf{k},-}$, and reaches a maximum value of $0.5$. This can be compared to the zero-temperature BCS value for the pairing amplitude evaluated on the FS, $\langle{c_{\textbf{k},\uparrow}}c_{-\textbf{k},\downarrow}\rangle = \frac{\Delta_{\textbf{k}}}{2\sqrt{\xi_{\textbf{k}}^2 + |\Delta_{\textbf{k}}|^2}} \xrightarrow[]{\text{FS}} \frac{1}{2}\frac{\Delta_{\textbf{k}}}{|\Delta_{\textbf{k}}|}$.

\section*{Appendix D: Three-orbital tight-binding model}

\renewcommand{\theequation}{D\arabic{equation}}
\setcounter{equation}{0}
\renewcommand\thefigure{D\arabic{figure}} 
\setcounter{figure}{0}
\setcounter{table}{0}
\renewcommand{\thetable}{D\arabic{table}}
\renewcommand{\theHfigure}{D\arabic{figure}}

The normal state Hamiltonian for the three-orbital model in Sec.~\ref{Five} of the main text is,
\begin{equation}
    H_{0} = \sum_{\textbf{k},\sigma,m}\xi^{m}_{\textbf{k}}c^{m\dagger}_{\textbf{k}\sigma}c^{m}_{\textbf{k}\sigma} + \sum_{\textbf{k},\sigma}t_{\textbf{k}}c^{yz\dagger}_{\textbf{k}\sigma}c^{xz}_{\textbf{k}\sigma} + \text{H.c.} + i\lambda\sum_{\textbf{k},l,m,n}\epsilon_{lmn}c^{l\dagger}_{\textbf{k}\sigma}c^{m}_{\textbf{k}\sigma'}\sigma^{n}_{\sigma\sigma'},
\end{equation}

\begin{table*}[htb!]
\centering
\setlength{\tabcolsep}{10pt}
\begin{center}
\begin{tabular}{c c c c c c c c} 
 \hline\hline
 $t_{1}$ & $t_{2}$ & $t_{3}$ & $t_{4}$ & $t_{5}$ & $\lambda$ & $\mu_{1}$ & $\mu_{2}$  \\ [0.5ex] 
 \hline
 0.45 & 0.05 & 0.5 & 0.2 & 0.025 & 0.085 & 0.531 & 0.631 \\  [0.5ex] 
 \hline\hline
\end{tabular}
\end{center}
 \caption{Tight-binding parameters for the three-orbital model from Ref.~\cite{lindquist2019distinct} used in Sec.~\hyperref[Five]{V} of the main text. All parameters are in units of $2t_{3}=1$.}
\label{table:tbparams}
\end{table*}

including the orbital dispersions $\xi^{m}_{\textbf{k}}$, hybridization between $(d_{yz},d_{xz})$ orbitals $t_{\textbf{k}}$, and atomic SOC $\lambda$. The SOC is written in terms of the completely antisymmetric tensor, $\epsilon_{lmn}$, with $(l,m,n) = (x,y,z)$ representing the $t_{2g}$ orbitals, $(yz,xz,xy)$. The orbital hybridization is, $t_{\textbf{k}} = -4t_{5}\sin{k_x}\sin{k_y}$, and the orbital dispersions are, $\xi_{\textbf{k}}^{yz/xz} = -2t_{1}\cos{k_{y/x}} - 2t_{2}\cos{k_{x/y}}-\mu_{1}-\mu, ~ \xi_{\textbf{k}}^{xy} = -2t_{3}(\cos{k_x} + \cos{k_y})-4t_{4}\cos{k_x}\cos{k_y}-\mu_{2}-\mu$. The parameters are given in Table~\ref{table:tbparams}, in units of $2t_{3}=1$, and the chemical potential, $\mu$, is adjusted to fix the electronic filling at $\frac{2}{3}$.

\twocolumngrid

\bibliography{main}

\begin{thebibliography}{73}%
\makeatletter
\providecommand \@ifxundefined [1]{%
 \@ifx{#1\undefined}
}%
\providecommand \@ifnum [1]{%
 \ifnum #1\expandafter \@firstoftwo
 \else \expandafter \@secondoftwo
 \fi
}%
\providecommand \@ifx [1]{%
 \ifx #1\expandafter \@firstoftwo
 \else \expandafter \@secondoftwo
 \fi
}%
\providecommand \natexlab [1]{#1}%
\providecommand \enquote  [1]{``#1''}%
\providecommand \bibnamefont  [1]{#1}%
\providecommand \bibfnamefont [1]{#1}%
\providecommand \citenamefont [1]{#1}%
\providecommand \href@noop [0]{\@secondoftwo}%
\providecommand \href [0]{\begingroup \@sanitize@url \@href}%
\providecommand \@href[1]{\@@startlink{#1}\@@href}%
\providecommand \@@href[1]{\endgroup#1\@@endlink}%
\providecommand \@sanitize@url [0]{\catcode `\\12\catcode `\$12\catcode `\&12\catcode `\#12\catcode `\^12\catcode `\_12\catcode `\%12\relax}%
\providecommand \@@startlink[1]{}%
\providecommand \@@endlink[0]{}%
\providecommand \url  [0]{\begingroup\@sanitize@url \@url }%
\providecommand \@url [1]{\endgroup\@href {#1}{\urlprefix }}%
\providecommand \urlprefix  [0]{URL }%
\providecommand \Eprint [0]{\href }%
\providecommand \doibase [0]{https://doi.org/}%
\providecommand \selectlanguage [0]{\@gobble}%
\providecommand \bibinfo  [0]{\@secondoftwo}%
\providecommand \bibfield  [0]{\@secondoftwo}%
\providecommand \translation [1]{[#1]}%
\providecommand \BibitemOpen [0]{}%
\providecommand \bibitemStop [0]{}%
\providecommand \bibitemNoStop [0]{.\EOS\space}%
\providecommand \EOS [0]{\spacefactor3000\relax}%
\providecommand \BibitemShut  [1]{\csname bibitem#1\endcsname}%
\let\auto@bib@innerbib\@empty
\bibitem [{\citenamefont {Fulde}\ and\ \citenamefont {Ferrell}(1964)}]{Fulde1964}%
  \BibitemOpen
  \bibfield  {author} {\bibinfo {author} {\bibfnamefont {P.}~\bibnamefont {Fulde}}\ and\ \bibinfo {author} {\bibfnamefont {R.~A.}\ \bibnamefont {Ferrell}},\ }\bibfield  {title} {\bibinfo {title} {Superconductivity in a {S}trong {S}pin-{E}xchange {F}ield},\ }\href {https://doi.org/10.1103/PhysRev.135.A550} {\bibfield  {journal} {\bibinfo  {journal} {Phys. Rev.}\ }\textbf {\bibinfo {volume} {135}},\ \bibinfo {pages} {A550} (\bibinfo {year} {1964})}\BibitemShut {NoStop}%
\bibitem [{\citenamefont {Larkin}\ and\ \citenamefont {Ovchinnikov}(1964)}]{Larkin1964}%
  \BibitemOpen
  \bibfield  {author} {\bibinfo {author} {\bibfnamefont {A.~I.}\ \bibnamefont {Larkin}}\ and\ \bibinfo {author} {\bibfnamefont {Y.~N.}\ \bibnamefont {Ovchinnikov}},\ }\bibfield  {title} {\bibinfo {title} {{Nonuniform state of superconductors}},\ }\href@noop {} {\bibfield  {journal} {\bibinfo  {journal} {Zh. Eksp. Teor. Fiz.}\ }\textbf {\bibinfo {volume} {47}},\ \bibinfo {pages} {1136} (\bibinfo {year} {1964})}\BibitemShut {NoStop}%
\bibitem [{\citenamefont {Agterberg}\ \emph {et~al.}(2020)\citenamefont {Agterberg}, \citenamefont {Davis}, \citenamefont {Edkins}, \citenamefont {Fradkin}, \citenamefont {Van~Harlingen}, \citenamefont {Kivelson}, \citenamefont {Lee}, \citenamefont {Radzihovsky}, \citenamefont {Tranquada},\ and\ \citenamefont {Wang}}]{agterberg2020}%
  \BibitemOpen
  \bibfield  {author} {\bibinfo {author} {\bibfnamefont {D.~F.}\ \bibnamefont {Agterberg}}, \bibinfo {author} {\bibfnamefont {J.~S.}\ \bibnamefont {Davis}}, \bibinfo {author} {\bibfnamefont {S.~D.}\ \bibnamefont {Edkins}}, \bibinfo {author} {\bibfnamefont {E.}~\bibnamefont {Fradkin}}, \bibinfo {author} {\bibfnamefont {D.~J.}\ \bibnamefont {Van~Harlingen}}, \bibinfo {author} {\bibfnamefont {S.~A.}\ \bibnamefont {Kivelson}}, \bibinfo {author} {\bibfnamefont {P.~A.}\ \bibnamefont {Lee}}, \bibinfo {author} {\bibfnamefont {L.}~\bibnamefont {Radzihovsky}}, \bibinfo {author} {\bibfnamefont {J.~M.}\ \bibnamefont {Tranquada}},\ and\ \bibinfo {author} {\bibfnamefont {Y.}~\bibnamefont {Wang}},\ }\bibfield  {title} {\bibinfo {title} {The {P}hysics of {P}air-{D}ensity {W}aves: Cuprate {S}uperconductors and {B}eyond},\ }\href {https://www.annualreviews.org/doi/10.1146/annurev-conmatphys-031119-050711} {\bibfield  {journal} {\bibinfo  {journal} {Annu. Rev. Condens. Matter Phys.}\ }\textbf {\bibinfo {volume} {11}},\ \bibinfo
  {pages} {231} (\bibinfo {year} {2020})}\BibitemShut {NoStop}%
\bibitem [{\citenamefont {Hamidian}\ \emph {et~al.}(2016)\citenamefont {Hamidian}, \citenamefont {Edkins}, \citenamefont {Joo}, \citenamefont {Kostin}, \citenamefont {Eisaki}, \citenamefont {Uchida}, \citenamefont {Lawler}, \citenamefont {Kim}, \citenamefont {Mackenzie}, \citenamefont {Fujita} \emph {et~al.}}]{hamidian_Nat2016}%
  \BibitemOpen
  \bibfield  {author} {\bibinfo {author} {\bibfnamefont {M.}~\bibnamefont {Hamidian}}, \bibinfo {author} {\bibfnamefont {S.}~\bibnamefont {Edkins}}, \bibinfo {author} {\bibfnamefont {S.~H.}\ \bibnamefont {Joo}}, \bibinfo {author} {\bibfnamefont {A.}~\bibnamefont {Kostin}}, \bibinfo {author} {\bibfnamefont {H.}~\bibnamefont {Eisaki}}, \bibinfo {author} {\bibfnamefont {S.}~\bibnamefont {Uchida}}, \bibinfo {author} {\bibfnamefont {M.}~\bibnamefont {Lawler}}, \bibinfo {author} {\bibfnamefont {E.-A.}\ \bibnamefont {Kim}}, \bibinfo {author} {\bibfnamefont {A.}~\bibnamefont {Mackenzie}}, \bibinfo {author} {\bibfnamefont {K.}~\bibnamefont {Fujita}}, \emph {et~al.},\ }\bibfield  {title} {\bibinfo {title} {Detection of a {C}ooper-pair density wave in {${\mathrm{Bi}}_{2}{\mathrm{Sr}}_{2}{\mathrm{CaCu}}_{2}{\mathrm{O}}_{8+x}$}},\ }\href {https://www.nature.com/articles/nature17411} {\bibfield  {journal} {\bibinfo  {journal} {Nature}\ }\textbf {\bibinfo {volume} {532}},\ \bibinfo {pages} {343} (\bibinfo {year}
  {2016})}\BibitemShut {NoStop}%
\bibitem [{\citenamefont {Chen}\ \emph {et~al.}(2021)\citenamefont {Chen}, \citenamefont {Yang}, \citenamefont {Hu}, \citenamefont {Zhao}, \citenamefont {Yuan}, \citenamefont {Xing}, \citenamefont {Qian}, \citenamefont {Huang}, \citenamefont {Li}, \citenamefont {Ye} \emph {et~al.}}]{chen_Nat2021}%
  \BibitemOpen
  \bibfield  {author} {\bibinfo {author} {\bibfnamefont {H.}~\bibnamefont {Chen}}, \bibinfo {author} {\bibfnamefont {H.}~\bibnamefont {Yang}}, \bibinfo {author} {\bibfnamefont {B.}~\bibnamefont {Hu}}, \bibinfo {author} {\bibfnamefont {Z.}~\bibnamefont {Zhao}}, \bibinfo {author} {\bibfnamefont {J.}~\bibnamefont {Yuan}}, \bibinfo {author} {\bibfnamefont {Y.}~\bibnamefont {Xing}}, \bibinfo {author} {\bibfnamefont {G.}~\bibnamefont {Qian}}, \bibinfo {author} {\bibfnamefont {Z.}~\bibnamefont {Huang}}, \bibinfo {author} {\bibfnamefont {G.}~\bibnamefont {Li}}, \bibinfo {author} {\bibfnamefont {Y.}~\bibnamefont {Ye}}, \emph {et~al.},\ }\bibfield  {title} {\bibinfo {title} {Roton pair density wave in a strong-coupling kagome superconductor},\ }\href {https://www.nature.com/articles/s41586-021-03983-5} {\bibfield  {journal} {\bibinfo  {journal} {Nature}\ }\textbf {\bibinfo {volume} {599}},\ \bibinfo {pages} {222} (\bibinfo {year} {2021})}\BibitemShut {NoStop}%
\bibitem [{\citenamefont {Liu}\ \emph {et~al.}(2021)\citenamefont {Liu}, \citenamefont {Chong}, \citenamefont {Sharma},\ and\ \citenamefont {Davis}}]{Liu_Science2021}%
  \BibitemOpen
  \bibfield  {author} {\bibinfo {author} {\bibfnamefont {X.}~\bibnamefont {Liu}}, \bibinfo {author} {\bibfnamefont {Y.~X.}\ \bibnamefont {Chong}}, \bibinfo {author} {\bibfnamefont {R.}~\bibnamefont {Sharma}},\ and\ \bibinfo {author} {\bibfnamefont {J.~C.~S.}\ \bibnamefont {Davis}},\ }\bibfield  {title} {\bibinfo {title} {Discovery of a {C}ooper-pair density wave state in a transition-metal dichalcogenide},\ }\href {https://doi.org/10.1126/science.abd4607} {\bibfield  {journal} {\bibinfo  {journal} {Science}\ }\textbf {\bibinfo {volume} {372}},\ \bibinfo {pages} {1447} (\bibinfo {year} {2021})}\BibitemShut {NoStop}%
\bibitem [{\citenamefont {Gu}\ \emph {et~al.}(2023)\citenamefont {Gu}, \citenamefont {Carroll}, \citenamefont {Wang}, \citenamefont {Ran}, \citenamefont {Broyles}, \citenamefont {Siddiquee}, \citenamefont {Butch}, \citenamefont {Saha}, \citenamefont {Paglione}, \citenamefont {Davis} \emph {et~al.}}]{gu_Nat2023}%
  \BibitemOpen
  \bibfield  {author} {\bibinfo {author} {\bibfnamefont {Q.}~\bibnamefont {Gu}}, \bibinfo {author} {\bibfnamefont {J.~P.}\ \bibnamefont {Carroll}}, \bibinfo {author} {\bibfnamefont {S.}~\bibnamefont {Wang}}, \bibinfo {author} {\bibfnamefont {S.}~\bibnamefont {Ran}}, \bibinfo {author} {\bibfnamefont {C.}~\bibnamefont {Broyles}}, \bibinfo {author} {\bibfnamefont {H.}~\bibnamefont {Siddiquee}}, \bibinfo {author} {\bibfnamefont {N.~P.}\ \bibnamefont {Butch}}, \bibinfo {author} {\bibfnamefont {S.~R.}\ \bibnamefont {Saha}}, \bibinfo {author} {\bibfnamefont {J.}~\bibnamefont {Paglione}}, \bibinfo {author} {\bibfnamefont {J.~S.}\ \bibnamefont {Davis}}, \emph {et~al.},\ }\bibfield  {title} {\bibinfo {title} {Detection of a pair density wave state in {$\mathrm{UTe}_{2}$}},\ }\href {https://www.nature.com/articles/s41586-023-05919-7} {\bibfield  {journal} {\bibinfo  {journal} {Nature}\ }\textbf {\bibinfo {volume} {618}},\ \bibinfo {pages} {921} (\bibinfo {year} {2023})}\BibitemShut {NoStop}%
\bibitem [{\citenamefont {Aishwarya}\ \emph {et~al.}(2023)\citenamefont {Aishwarya}, \citenamefont {May-Mann}, \citenamefont {Raghavan}, \citenamefont {Nie}, \citenamefont {Romanelli}, \citenamefont {Ran}, \citenamefont {Saha}, \citenamefont {Paglione}, \citenamefont {Butch}, \citenamefont {Fradkin} \emph {et~al.}}]{aishwarya_Nat2023}%
  \BibitemOpen
  \bibfield  {author} {\bibinfo {author} {\bibfnamefont {A.}~\bibnamefont {Aishwarya}}, \bibinfo {author} {\bibfnamefont {J.}~\bibnamefont {May-Mann}}, \bibinfo {author} {\bibfnamefont {A.}~\bibnamefont {Raghavan}}, \bibinfo {author} {\bibfnamefont {L.}~\bibnamefont {Nie}}, \bibinfo {author} {\bibfnamefont {M.}~\bibnamefont {Romanelli}}, \bibinfo {author} {\bibfnamefont {S.}~\bibnamefont {Ran}}, \bibinfo {author} {\bibfnamefont {S.~R.}\ \bibnamefont {Saha}}, \bibinfo {author} {\bibfnamefont {J.}~\bibnamefont {Paglione}}, \bibinfo {author} {\bibfnamefont {N.~P.}\ \bibnamefont {Butch}}, \bibinfo {author} {\bibfnamefont {E.}~\bibnamefont {Fradkin}}, \emph {et~al.},\ }\bibfield  {title} {\bibinfo {title} {Magnetic-field-sensitive charge density waves in the superconductor {$\mathrm{UTe}_{2}$}},\ }\href {https://www.nature.com/articles/s41586-023-06005-8} {\bibfield  {journal} {\bibinfo  {journal} {Nature}\ }\textbf {\bibinfo {volume} {618}},\ \bibinfo {pages} {928} (\bibinfo {year} {2023})}\BibitemShut {NoStop}%
\bibitem [{\citenamefont {Liu}\ \emph {et~al.}(2023)\citenamefont {Liu}, \citenamefont {Wei}, \citenamefont {He}, \citenamefont {Zhang}, \citenamefont {Wang},\ and\ \citenamefont {Wang}}]{liu_Nature2023}%
  \BibitemOpen
  \bibfield  {author} {\bibinfo {author} {\bibfnamefont {Y.}~\bibnamefont {Liu}}, \bibinfo {author} {\bibfnamefont {T.}~\bibnamefont {Wei}}, \bibinfo {author} {\bibfnamefont {G.}~\bibnamefont {He}}, \bibinfo {author} {\bibfnamefont {Y.}~\bibnamefont {Zhang}}, \bibinfo {author} {\bibfnamefont {Z.}~\bibnamefont {Wang}},\ and\ \bibinfo {author} {\bibfnamefont {J.}~\bibnamefont {Wang}},\ }\bibfield  {title} {\bibinfo {title} {Pair density wave state in a monolayer high-{$T_{c}$} iron-based superconductor},\ }\href {https://www.nature.com/articles/s41586-023-06072-x} {\bibfield  {journal} {\bibinfo  {journal} {Nature}\ }\textbf {\bibinfo {volume} {618}},\ \bibinfo {pages} {934} (\bibinfo {year} {2023})}\BibitemShut {NoStop}%
\bibitem [{\citenamefont {Zhao}\ \emph {et~al.}(2023)\citenamefont {Zhao}, \citenamefont {Blackwell}, \citenamefont {Thinel}, \citenamefont {Handa}, \citenamefont {Ishida}, \citenamefont {Zhu}, \citenamefont {Iyo}, \citenamefont {Eisaki}, \citenamefont {Pasupathy},\ and\ \citenamefont {Fujita}}]{zhao_Nature2023}%
  \BibitemOpen
  \bibfield  {author} {\bibinfo {author} {\bibfnamefont {H.}~\bibnamefont {Zhao}}, \bibinfo {author} {\bibfnamefont {R.}~\bibnamefont {Blackwell}}, \bibinfo {author} {\bibfnamefont {M.}~\bibnamefont {Thinel}}, \bibinfo {author} {\bibfnamefont {T.}~\bibnamefont {Handa}}, \bibinfo {author} {\bibfnamefont {S.}~\bibnamefont {Ishida}}, \bibinfo {author} {\bibfnamefont {X.}~\bibnamefont {Zhu}}, \bibinfo {author} {\bibfnamefont {A.}~\bibnamefont {Iyo}}, \bibinfo {author} {\bibfnamefont {H.}~\bibnamefont {Eisaki}}, \bibinfo {author} {\bibfnamefont {A.~N.}\ \bibnamefont {Pasupathy}},\ and\ \bibinfo {author} {\bibfnamefont {K.}~\bibnamefont {Fujita}},\ }\bibfield  {title} {\bibinfo {title} {Smectic pair-density-wave order in {$\mathrm{EuRbFe}_{4}\mathrm{As}_{4}$}},\ }\href {https://www.nature.com/articles/s41586-023-06103-7} {\bibfield  {journal} {\bibinfo  {journal} {Nature}\ }\textbf {\bibinfo {volume} {618}},\ \bibinfo {pages} {940} (\bibinfo {year} {2023})}\BibitemShut {NoStop}%
\bibitem [{\citenamefont {Cho}\ \emph {et~al.}(2011)\citenamefont {Cho}, \citenamefont {Kim}, \citenamefont {Tanatar}, \citenamefont {Song}, \citenamefont {Kwon}, \citenamefont {Coniglio}, \citenamefont {Agosta}, \citenamefont {Gurevich},\ and\ \citenamefont {Prozorov}}]{Cho_PRB2011}%
  \BibitemOpen
  \bibfield  {author} {\bibinfo {author} {\bibfnamefont {K.}~\bibnamefont {Cho}}, \bibinfo {author} {\bibfnamefont {H.}~\bibnamefont {Kim}}, \bibinfo {author} {\bibfnamefont {M.~A.}\ \bibnamefont {Tanatar}}, \bibinfo {author} {\bibfnamefont {Y.~J.}\ \bibnamefont {Song}}, \bibinfo {author} {\bibfnamefont {Y.~S.}\ \bibnamefont {Kwon}}, \bibinfo {author} {\bibfnamefont {W.~A.}\ \bibnamefont {Coniglio}}, \bibinfo {author} {\bibfnamefont {C.~C.}\ \bibnamefont {Agosta}}, \bibinfo {author} {\bibfnamefont {A.}~\bibnamefont {Gurevich}},\ and\ \bibinfo {author} {\bibfnamefont {R.}~\bibnamefont {Prozorov}},\ }\bibfield  {title} {\bibinfo {title} {Anisotropic upper critical field and possible {F}ulde-{F}errel-{L}arkin-{O}vchinnikov state in the stoichiometric pnictide superconductor {$\mathrm{LiFeAs}$}},\ }\href {https://doi.org/10.1103/PhysRevB.83.060502} {\bibfield  {journal} {\bibinfo  {journal} {Phys. Rev. B}\ }\textbf {\bibinfo {volume} {83}},\ \bibinfo {pages} {060502} (\bibinfo {year} {2011})}\BibitemShut
  {NoStop}%
\bibitem [{\citenamefont {Tarantini}\ \emph {et~al.}(2011)\citenamefont {Tarantini}, \citenamefont {Gurevich}, \citenamefont {Jaroszynski}, \citenamefont {Balakirev}, \citenamefont {Bellingeri}, \citenamefont {Pallecchi}, \citenamefont {Ferdeghini}, \citenamefont {Shen}, \citenamefont {Wen},\ and\ \citenamefont {Larbalestier}}]{Tarantini_PRB2011}%
  \BibitemOpen
  \bibfield  {author} {\bibinfo {author} {\bibfnamefont {C.}~\bibnamefont {Tarantini}}, \bibinfo {author} {\bibfnamefont {A.}~\bibnamefont {Gurevich}}, \bibinfo {author} {\bibfnamefont {J.}~\bibnamefont {Jaroszynski}}, \bibinfo {author} {\bibfnamefont {F.}~\bibnamefont {Balakirev}}, \bibinfo {author} {\bibfnamefont {E.}~\bibnamefont {Bellingeri}}, \bibinfo {author} {\bibfnamefont {I.}~\bibnamefont {Pallecchi}}, \bibinfo {author} {\bibfnamefont {C.}~\bibnamefont {Ferdeghini}}, \bibinfo {author} {\bibfnamefont {B.}~\bibnamefont {Shen}}, \bibinfo {author} {\bibfnamefont {H.~H.}\ \bibnamefont {Wen}},\ and\ \bibinfo {author} {\bibfnamefont {D.~C.}\ \bibnamefont {Larbalestier}},\ }\bibfield  {title} {\bibinfo {title} {Significant enhancement of upper critical fields by doping and strain in iron-based superconductors},\ }\href {https://doi.org/10.1103/PhysRevB.84.184522} {\bibfield  {journal} {\bibinfo  {journal} {Phys. Rev. B}\ }\textbf {\bibinfo {volume} {84}},\ \bibinfo {pages} {184522} (\bibinfo {year}
  {2011})}\BibitemShut {NoStop}%
\bibitem [{\citenamefont {Kasahara}\ \emph {et~al.}(2020)\citenamefont {Kasahara}, \citenamefont {Sato}, \citenamefont {Licciardello}, \citenamefont {\ifmmode~\check{C}\else \v{C}\fi{}ulo}, \citenamefont {Arsenijevi\ifmmode~\acute{c}\else \'{c}\fi{}}, \citenamefont {Ottenbros}, \citenamefont {Tominaga}, \citenamefont {B\"oker}, \citenamefont {Eremin}, \citenamefont {Shibauchi}, \citenamefont {Wosnitza}, \citenamefont {Hussey},\ and\ \citenamefont {Matsuda}}]{Kasahara_PRL2020}%
  \BibitemOpen
  \bibfield  {author} {\bibinfo {author} {\bibfnamefont {S.}~\bibnamefont {Kasahara}}, \bibinfo {author} {\bibfnamefont {Y.}~\bibnamefont {Sato}}, \bibinfo {author} {\bibfnamefont {S.}~\bibnamefont {Licciardello}}, \bibinfo {author} {\bibfnamefont {M.}~\bibnamefont {\ifmmode~\check{C}\else \v{C}\fi{}ulo}}, \bibinfo {author} {\bibfnamefont {S.}~\bibnamefont {Arsenijevi\ifmmode~\acute{c}\else \'{c}\fi{}}}, \bibinfo {author} {\bibfnamefont {T.}~\bibnamefont {Ottenbros}}, \bibinfo {author} {\bibfnamefont {T.}~\bibnamefont {Tominaga}}, \bibinfo {author} {\bibfnamefont {J.}~\bibnamefont {B\"oker}}, \bibinfo {author} {\bibfnamefont {I.}~\bibnamefont {Eremin}}, \bibinfo {author} {\bibfnamefont {T.}~\bibnamefont {Shibauchi}}, \bibinfo {author} {\bibfnamefont {J.}~\bibnamefont {Wosnitza}}, \bibinfo {author} {\bibfnamefont {N.~E.}\ \bibnamefont {Hussey}},\ and\ \bibinfo {author} {\bibfnamefont {Y.}~\bibnamefont {Matsuda}},\ }\bibfield  {title} {\bibinfo {title} {Evidence for an {F}ulde-{F}errell-{L}arkin-{O}vchinnikov
  {S}tate with {S}egmented {V}ortices in the {BCS}-{BEC}-{C}rossover {S}uperconductor {$\mathrm{FeSe}$}},\ }\href {https://doi.org/10.1103/PhysRevLett.124.107001} {\bibfield  {journal} {\bibinfo  {journal} {Phys. Rev. Lett.}\ }\textbf {\bibinfo {volume} {124}},\ \bibinfo {pages} {107001} (\bibinfo {year} {2020})}\BibitemShut {NoStop}%
\bibitem [{\citenamefont {Kinjo}\ \emph {et~al.}(2022)\citenamefont {Kinjo}, \citenamefont {Manago}, \citenamefont {Kitagawa}, \citenamefont {Mao}, \citenamefont {Yonezawa}, \citenamefont {Maeno},\ and\ \citenamefont {Ishida}}]{Kinjo_Science2022}%
  \BibitemOpen
  \bibfield  {author} {\bibinfo {author} {\bibfnamefont {K.}~\bibnamefont {Kinjo}}, \bibinfo {author} {\bibfnamefont {M.}~\bibnamefont {Manago}}, \bibinfo {author} {\bibfnamefont {S.}~\bibnamefont {Kitagawa}}, \bibinfo {author} {\bibfnamefont {Z.~Q.}\ \bibnamefont {Mao}}, \bibinfo {author} {\bibfnamefont {S.}~\bibnamefont {Yonezawa}}, \bibinfo {author} {\bibfnamefont {Y.}~\bibnamefont {Maeno}},\ and\ \bibinfo {author} {\bibfnamefont {K.}~\bibnamefont {Ishida}},\ }\bibfield  {title} {\bibinfo {title} {Superconducting spin smecticity evidencing the {F}ulde-{F}errell-{L}arkin-{O}vchinnikov state in {${\mathrm{Sr}}_{2}{\mathrm{RuO}}_{4}$}},\ }\href {https://doi.org/10.1126/science.abb0332} {\bibfield  {journal} {\bibinfo  {journal} {Science}\ }\textbf {\bibinfo {volume} {376}},\ \bibinfo {pages} {397} (\bibinfo {year} {2022})}\BibitemShut {NoStop}%
\bibitem [{\citenamefont {Spa\l{}ek}(2001)}]{SpalekPRB2001}%
  \BibitemOpen
  \bibfield  {author} {\bibinfo {author} {\bibfnamefont {J.}~\bibnamefont {Spa\l{}ek}},\ }\bibfield  {title} {\bibinfo {title} {Spin-triplet superconducting pairing due to local {H}und's rule and {D}irac exchange},\ }\href {https://doi.org/10.1103/PhysRevB.63.104513} {\bibfield  {journal} {\bibinfo  {journal} {Phys. Rev. B}\ }\textbf {\bibinfo {volume} {63}},\ \bibinfo {pages} {104513} (\bibinfo {year} {2001})}\BibitemShut {NoStop}%
\bibitem [{\citenamefont {Dai}\ \emph {et~al.}(2008)\citenamefont {Dai}, \citenamefont {Fang}, \citenamefont {Zhou},\ and\ \citenamefont {Zhang}}]{Dai2008PRL}%
  \BibitemOpen
  \bibfield  {author} {\bibinfo {author} {\bibfnamefont {X.}~\bibnamefont {Dai}}, \bibinfo {author} {\bibfnamefont {Z.}~\bibnamefont {Fang}}, \bibinfo {author} {\bibfnamefont {Y.}~\bibnamefont {Zhou}},\ and\ \bibinfo {author} {\bibfnamefont {F.-C.}\ \bibnamefont {Zhang}},\ }\bibfield  {title} {\bibinfo {title} {Even {P}arity, {O}rbital {S}inglet, and {S}pin {T}riplet {P}airing for {S}uperconducting {${\mathrm{LaFeAsO}}_{1\ensuremath{-}x}{\mathrm{F}}_{x}$}},\ }\href {https://doi.org/10.1103/PhysRevLett.101.057008} {\bibfield  {journal} {\bibinfo  {journal} {Phys. Rev. Lett.}\ }\textbf {\bibinfo {volume} {101}},\ \bibinfo {pages} {057008} (\bibinfo {year} {2008})}\BibitemShut {NoStop}%
\bibitem [{\citenamefont {Puetter}\ and\ \citenamefont {Kee}(2012)}]{Puetter2012EPL}%
  \BibitemOpen
  \bibfield  {author} {\bibinfo {author} {\bibfnamefont {C.~M.}\ \bibnamefont {Puetter}}\ and\ \bibinfo {author} {\bibfnamefont {H.-Y.}\ \bibnamefont {Kee}},\ }\bibfield  {title} {\bibinfo {title} {Identifying spin-triplet pairing in spin-orbit coupled multi-band superconductors},\ }\href {https://doi.org/10.1209/0295-5075/98/27010} {\bibfield  {journal} {\bibinfo  {journal} {Europhys. Lett.}\ }\textbf {\bibinfo {volume} {98}},\ \bibinfo {pages} {27010} (\bibinfo {year} {2012})}\BibitemShut {NoStop}%
\bibitem [{\citenamefont {Hoshino}\ and\ \citenamefont {Werner}(2015)}]{Hoshino2015PRL}%
  \BibitemOpen
  \bibfield  {author} {\bibinfo {author} {\bibfnamefont {S.}~\bibnamefont {Hoshino}}\ and\ \bibinfo {author} {\bibfnamefont {P.}~\bibnamefont {Werner}},\ }\bibfield  {title} {\bibinfo {title} {Superconductivity from {E}merging {M}agnetic {M}oments},\ }\href {https://doi.org/10.1103/PhysRevLett.115.247001} {\bibfield  {journal} {\bibinfo  {journal} {Phys. Rev. Lett.}\ }\textbf {\bibinfo {volume} {115}},\ \bibinfo {pages} {247001} (\bibinfo {year} {2015})}\BibitemShut {NoStop}%
\bibitem [{\citenamefont {Hoshino}\ and\ \citenamefont {Werner}(2016)}]{Hoshino2016PRB}%
  \BibitemOpen
  \bibfield  {author} {\bibinfo {author} {\bibfnamefont {S.}~\bibnamefont {Hoshino}}\ and\ \bibinfo {author} {\bibfnamefont {P.}~\bibnamefont {Werner}},\ }\bibfield  {title} {\bibinfo {title} {Electronic orders in multiorbital {H}ubbard models with lifted orbital degeneracy},\ }\href {https://doi.org/10.1103/PhysRevB.93.155161} {\bibfield  {journal} {\bibinfo  {journal} {Phys. Rev. B}\ }\textbf {\bibinfo {volume} {93}},\ \bibinfo {pages} {155161} (\bibinfo {year} {2016})}\BibitemShut {NoStop}%
\bibitem [{\citenamefont {Vafek}\ and\ \citenamefont {Chubukov}(2017)}]{vafek2017hund}%
  \BibitemOpen
  \bibfield  {author} {\bibinfo {author} {\bibfnamefont {O.}~\bibnamefont {Vafek}}\ and\ \bibinfo {author} {\bibfnamefont {A.~V.}\ \bibnamefont {Chubukov}},\ }\bibfield  {title} {\bibinfo {title} {Hund {I}nteraction, {S}pin-{O}rbit {C}oupling, and the {M}echanism of {S}uperconductivity in {S}trongly {H}ole-{D}oped {I}ron {P}nictides},\ }\href {https://link.aps.org/doi/10.1103/PhysRevLett.118.087003} {\bibfield  {journal} {\bibinfo  {journal} {Phys. Rev. Lett.}\ }\textbf {\bibinfo {volume} {118}},\ \bibinfo {pages} {087003} (\bibinfo {year} {2017})}\BibitemShut {NoStop}%
\bibitem [{\citenamefont {Gingras}\ \emph {et~al.}(2019)\citenamefont {Gingras}, \citenamefont {Nourafkan}, \citenamefont {Tremblay},\ and\ \citenamefont {C\^ot\'e}}]{Gingras2019PRL}%
  \BibitemOpen
  \bibfield  {author} {\bibinfo {author} {\bibfnamefont {O.}~\bibnamefont {Gingras}}, \bibinfo {author} {\bibfnamefont {R.}~\bibnamefont {Nourafkan}}, \bibinfo {author} {\bibfnamefont {A.-M.~S.}\ \bibnamefont {Tremblay}},\ and\ \bibinfo {author} {\bibfnamefont {M.}~\bibnamefont {C\^ot\'e}},\ }\bibfield  {title} {\bibinfo {title} {Superconducting {S}ymmetries of {${\mathrm{Sr}}_{2}{\mathrm{RuO}}_{4}$} from {F}irst-{P}rinciples {E}lectronic {S}tructure},\ }\href {https://doi.org/10.1103/PhysRevLett.123.217005} {\bibfield  {journal} {\bibinfo  {journal} {Phys. Rev. Lett.}\ }\textbf {\bibinfo {volume} {123}},\ \bibinfo {pages} {217005} (\bibinfo {year} {2019})}\BibitemShut {NoStop}%
\bibitem [{\citenamefont {Lindquist}\ and\ \citenamefont {Kee}(2020)}]{lindquist2019distinct}%
  \BibitemOpen
  \bibfield  {author} {\bibinfo {author} {\bibfnamefont {A.~W.}\ \bibnamefont {Lindquist}}\ and\ \bibinfo {author} {\bibfnamefont {H.-Y.}\ \bibnamefont {Kee}},\ }\bibfield  {title} {\bibinfo {title} {Distinct reduction of {K}night shift in superconducting state of {${\mathrm{Sr}}_{2}{\mathrm{RuO}}_{4}$} under uniaxial strain},\ }\href {https://doi.org/10.1103/PhysRevResearch.2.032055} {\bibfield  {journal} {\bibinfo  {journal} {Phys. Rev. Res.}\ }\textbf {\bibinfo {volume} {2}},\ \bibinfo {pages} {032055} (\bibinfo {year} {2020})}\BibitemShut {NoStop}%
\bibitem [{\citenamefont {Coleman}\ \emph {et~al.}(2020)\citenamefont {Coleman}, \citenamefont {Komijani},\ and\ \citenamefont {K\"onig}}]{Coleman_PRL2020}%
  \BibitemOpen
  \bibfield  {author} {\bibinfo {author} {\bibfnamefont {P.}~\bibnamefont {Coleman}}, \bibinfo {author} {\bibfnamefont {Y.}~\bibnamefont {Komijani}},\ and\ \bibinfo {author} {\bibfnamefont {E.~J.}\ \bibnamefont {K\"onig}},\ }\bibfield  {title} {\bibinfo {title} {Triplet {R}esonating {V}alence {B}ond {S}tate and {S}uperconductivity in {H}und's {M}etals},\ }\href {https://doi.org/10.1103/PhysRevLett.125.077001} {\bibfield  {journal} {\bibinfo  {journal} {Phys. Rev. Lett.}\ }\textbf {\bibinfo {volume} {125}},\ \bibinfo {pages} {077001} (\bibinfo {year} {2020})}\BibitemShut {NoStop}%
\bibitem [{\citenamefont {Suh}\ \emph {et~al.}(2020)\citenamefont {Suh}, \citenamefont {Menke}, \citenamefont {Brydon}, \citenamefont {Timm}, \citenamefont {Ramires},\ and\ \citenamefont {Agterberg}}]{Suh2019}%
  \BibitemOpen
  \bibfield  {author} {\bibinfo {author} {\bibfnamefont {H.~G.}\ \bibnamefont {Suh}}, \bibinfo {author} {\bibfnamefont {H.}~\bibnamefont {Menke}}, \bibinfo {author} {\bibfnamefont {P.~M.~R.}\ \bibnamefont {Brydon}}, \bibinfo {author} {\bibfnamefont {C.}~\bibnamefont {Timm}}, \bibinfo {author} {\bibfnamefont {A.}~\bibnamefont {Ramires}},\ and\ \bibinfo {author} {\bibfnamefont {D.~F.}\ \bibnamefont {Agterberg}},\ }\bibfield  {title} {\bibinfo {title} {Stabilizing even-parity chiral superconductivity in {${\mathrm{Sr}}_{2}{\mathrm{RuO}}_{4}$}},\ }\href {https://link.aps.org/doi/10.1103/PhysRevResearch.2.032023} {\bibfield  {journal} {\bibinfo  {journal} {Phys. Rev. Res.}\ }\textbf {\bibinfo {volume} {2}},\ \bibinfo {pages} {032023} (\bibinfo {year} {2020})}\BibitemShut {NoStop}%
\bibitem [{\citenamefont {Clepkens}\ \emph {et~al.}(2021{\natexlab{a}})\citenamefont {Clepkens}, \citenamefont {Lindquist}, \citenamefont {Liu},\ and\ \citenamefont {Kee}}]{clepkens2021higher}%
  \BibitemOpen
  \bibfield  {author} {\bibinfo {author} {\bibfnamefont {J.}~\bibnamefont {Clepkens}}, \bibinfo {author} {\bibfnamefont {A.~W.}\ \bibnamefont {Lindquist}}, \bibinfo {author} {\bibfnamefont {X.}~\bibnamefont {Liu}},\ and\ \bibinfo {author} {\bibfnamefont {H.-Y.}\ \bibnamefont {Kee}},\ }\bibfield  {title} {\bibinfo {title} {Higher angular momentum pairings in interorbital shadowed-triplet superconductors: Application to {${\mathrm{Sr}}_{2}{\mathrm{RuO}}_{4}$}},\ }\href {https://doi.org/10.1103/PhysRevB.104.104512} {\bibfield  {journal} {\bibinfo  {journal} {Phys. Rev. B}\ }\textbf {\bibinfo {volume} {104}},\ \bibinfo {pages} {104512} (\bibinfo {year} {2021}{\natexlab{a}})}\BibitemShut {NoStop}%
\bibitem [{\citenamefont {Berg}\ \emph {et~al.}(2009)\citenamefont {Berg}, \citenamefont {Fradkin},\ and\ \citenamefont {Kivelson}}]{Berg_PRB2009}%
  \BibitemOpen
  \bibfield  {author} {\bibinfo {author} {\bibfnamefont {E.}~\bibnamefont {Berg}}, \bibinfo {author} {\bibfnamefont {E.}~\bibnamefont {Fradkin}},\ and\ \bibinfo {author} {\bibfnamefont {S.~A.}\ \bibnamefont {Kivelson}},\ }\bibfield  {title} {\bibinfo {title} {Theory of the striped superconductor},\ }\href {https://doi.org/10.1103/PhysRevB.79.064515} {\bibfield  {journal} {\bibinfo  {journal} {Phys. Rev. B}\ }\textbf {\bibinfo {volume} {79}},\ \bibinfo {pages} {064515} (\bibinfo {year} {2009})}\BibitemShut {NoStop}%
\bibitem [{\citenamefont {Jiang}\ and\ \citenamefont {Yao}(2023)}]{jiang_2023}%
  \BibitemOpen
  \bibfield  {author} {\bibinfo {author} {\bibfnamefont {Y.-F.}\ \bibnamefont {Jiang}}\ and\ \bibinfo {author} {\bibfnamefont {H.}~\bibnamefont {Yao}},\ }\href@noop {} {\bibinfo {title} {Pair density wave superconductivity: a microscopic model in two dimensions}} (\bibinfo {year} {2023}),\ \Eprint {https://arxiv.org/abs/2308.08609} {arXiv:2308.08609 [cond-mat.supr-con]} \BibitemShut {NoStop}%
\bibitem [{\citenamefont {Jian}\ \emph {et~al.}(2015)\citenamefont {Jian}, \citenamefont {Jiang},\ and\ \citenamefont {Yao}}]{Jian_PRL2015}%
  \BibitemOpen
  \bibfield  {author} {\bibinfo {author} {\bibfnamefont {S.-K.}\ \bibnamefont {Jian}}, \bibinfo {author} {\bibfnamefont {Y.-F.}\ \bibnamefont {Jiang}},\ and\ \bibinfo {author} {\bibfnamefont {H.}~\bibnamefont {Yao}},\ }\bibfield  {title} {\bibinfo {title} {Emergent {S}pacetime {S}upersymmetry in 3{D} {W}eyl {S}emimetals and 2{D} {D}irac {S}emimetals},\ }\href {https://doi.org/10.1103/PhysRevLett.114.237001} {\bibfield  {journal} {\bibinfo  {journal} {Phys. Rev. Lett.}\ }\textbf {\bibinfo {volume} {114}},\ \bibinfo {pages} {237001} (\bibinfo {year} {2015})}\BibitemShut {NoStop}%
\bibitem [{\citenamefont {Jiang}\ and\ \citenamefont {Barlas}(2023)}]{Jiang_PRL2023}%
  \BibitemOpen
  \bibfield  {author} {\bibinfo {author} {\bibfnamefont {G.}~\bibnamefont {Jiang}}\ and\ \bibinfo {author} {\bibfnamefont {Y.}~\bibnamefont {Barlas}},\ }\bibfield  {title} {\bibinfo {title} {Pair {D}ensity {W}aves from {L}ocal {B}and {G}eometry},\ }\href {https://doi.org/10.1103/PhysRevLett.131.016002} {\bibfield  {journal} {\bibinfo  {journal} {Phys. Rev. Lett.}\ }\textbf {\bibinfo {volume} {131}},\ \bibinfo {pages} {016002} (\bibinfo {year} {2023})}\BibitemShut {NoStop}%
\bibitem [{\citenamefont {Chen}\ and\ \citenamefont {Huang}(2023)}]{chen_SCP2023}%
  \BibitemOpen
  \bibfield  {author} {\bibinfo {author} {\bibfnamefont {W.}~\bibnamefont {Chen}}\ and\ \bibinfo {author} {\bibfnamefont {W.}~\bibnamefont {Huang}},\ }\bibfield  {title} {\bibinfo {title} {Pair density wave facilitated by {B}loch quantum geometry in nearly flat band multiorbital superconductors},\ }\href {https://link.springer.com/article/10.1007/s11433-023-2122-4} {\bibfield  {journal} {\bibinfo  {journal} {Science China Physics, Mechanics \& Astronomy}\ }\textbf {\bibinfo {volume} {66}},\ \bibinfo {pages} {287212} (\bibinfo {year} {2023})}\BibitemShut {NoStop}%
\bibitem [{\citenamefont {Han}\ and\ \citenamefont {Kivelson}(2022)}]{Han_PRB2022}%
  \BibitemOpen
  \bibfield  {author} {\bibinfo {author} {\bibfnamefont {Z.}~\bibnamefont {Han}}\ and\ \bibinfo {author} {\bibfnamefont {S.~A.}\ \bibnamefont {Kivelson}},\ }\bibfield  {title} {\bibinfo {title} {Pair density wave and reentrant superconducting tendencies originating from valley polarization},\ }\href {https://doi.org/10.1103/PhysRevB.105.L100509} {\bibfield  {journal} {\bibinfo  {journal} {Phys. Rev. B}\ }\textbf {\bibinfo {volume} {105}},\ \bibinfo {pages} {L100509} (\bibinfo {year} {2022})}\BibitemShut {NoStop}%
\bibitem [{\citenamefont {Chakraborty}\ and\ \citenamefont {Black-Schaffer}(2023)}]{chakraborty_2023}%
  \BibitemOpen
  \bibfield  {author} {\bibinfo {author} {\bibfnamefont {D.}~\bibnamefont {Chakraborty}}\ and\ \bibinfo {author} {\bibfnamefont {A.~M.}\ \bibnamefont {Black-Schaffer}},\ }\href@noop {} {\bibinfo {title} {Zero-field finite-momentum and field-induced superconductivity in altermagnets}} (\bibinfo {year} {2023}),\ \Eprint {https://arxiv.org/abs/2309.14427} {arXiv:2309.14427 [cond-mat.supr-con]} \BibitemShut {NoStop}%
\bibitem [{\citenamefont {Kang}\ and\ \citenamefont {Te\ifmmode \check{s}\else \v{s}\fi{}anovi\ifmmode~\acute{c}\else \'{c}\fi{}}(2011)}]{Kang_PRB2011}%
  \BibitemOpen
  \bibfield  {author} {\bibinfo {author} {\bibfnamefont {J.}~\bibnamefont {Kang}}\ and\ \bibinfo {author} {\bibfnamefont {Z.}~\bibnamefont {Te\ifmmode \check{s}\else \v{s}\fi{}anovi\ifmmode~\acute{c}\else \'{c}\fi{}}},\ }\bibfield  {title} {\bibinfo {title} {Theory of the valley-density wave and hidden order in iron pnictides},\ }\href {https://doi.org/10.1103/PhysRevB.83.020505} {\bibfield  {journal} {\bibinfo  {journal} {Phys. Rev. B}\ }\textbf {\bibinfo {volume} {83}},\ \bibinfo {pages} {020505} (\bibinfo {year} {2011})}\BibitemShut {NoStop}%
\bibitem [{\citenamefont {Lee}\ \emph {et~al.}(2007)\citenamefont {Lee}, \citenamefont {Lee},\ and\ \citenamefont {Senthil}}]{Lee_PRL2007}%
  \BibitemOpen
  \bibfield  {author} {\bibinfo {author} {\bibfnamefont {S.-S.}\ \bibnamefont {Lee}}, \bibinfo {author} {\bibfnamefont {P.~A.}\ \bibnamefont {Lee}},\ and\ \bibinfo {author} {\bibfnamefont {T.}~\bibnamefont {Senthil}},\ }\bibfield  {title} {\bibinfo {title} {Amperean {P}airing {I}nstability in the {U}(1) {S}pin {L}iquid {S}tate with {F}ermi {S}urface and {A}pplication to $\ensuremath{\kappa}\mathrm{\text{\ensuremath{-}}}(\mathrm{BEDT}\mathrm{\text{\ensuremath{-}}}\mathrm{TTF}{)}_{2}{{\mathrm{Cu}}_{2}(\mathrm{CN}{)}_{3}}$},\ }\href {https://doi.org/10.1103/PhysRevLett.98.067006} {\bibfield  {journal} {\bibinfo  {journal} {Phys. Rev. Lett.}\ }\textbf {\bibinfo {volume} {98}},\ \bibinfo {pages} {067006} (\bibinfo {year} {2007})}\BibitemShut {NoStop}%
\bibitem [{\citenamefont {Soto-Garrido}\ \emph {et~al.}(2015)\citenamefont {Soto-Garrido}, \citenamefont {Cho},\ and\ \citenamefont {Fradkin}}]{SotoGarrido_PRB2015}%
  \BibitemOpen
  \bibfield  {author} {\bibinfo {author} {\bibfnamefont {R.}~\bibnamefont {Soto-Garrido}}, \bibinfo {author} {\bibfnamefont {G.~Y.}\ \bibnamefont {Cho}},\ and\ \bibinfo {author} {\bibfnamefont {E.}~\bibnamefont {Fradkin}},\ }\bibfield  {title} {\bibinfo {title} {Quasi-one-dimensional pair density wave superconducting state},\ }\href {https://doi.org/10.1103/PhysRevB.91.195102} {\bibfield  {journal} {\bibinfo  {journal} {Phys. Rev. B}\ }\textbf {\bibinfo {volume} {91}},\ \bibinfo {pages} {195102} (\bibinfo {year} {2015})}\BibitemShut {NoStop}%
\bibitem [{\citenamefont {Loder}\ \emph {et~al.}(2011)\citenamefont {Loder}, \citenamefont {Graser}, \citenamefont {Kampf},\ and\ \citenamefont {Kopp}}]{Florian_PRL2011}%
  \BibitemOpen
  \bibfield  {author} {\bibinfo {author} {\bibfnamefont {F.}~\bibnamefont {Loder}}, \bibinfo {author} {\bibfnamefont {S.}~\bibnamefont {Graser}}, \bibinfo {author} {\bibfnamefont {A.~P.}\ \bibnamefont {Kampf}},\ and\ \bibinfo {author} {\bibfnamefont {T.}~\bibnamefont {Kopp}},\ }\bibfield  {title} {\bibinfo {title} {Mean-{F}ield {P}airing {T}heory for the {C}harge-{S}tripe {P}hase of {H}igh-{T}emperature {C}uprate {S}uperconductors},\ }\href {https://doi.org/10.1103/PhysRevLett.107.187001} {\bibfield  {journal} {\bibinfo  {journal} {Phys. Rev. Lett.}\ }\textbf {\bibinfo {volume} {107}},\ \bibinfo {pages} {187001} (\bibinfo {year} {2011})}\BibitemShut {NoStop}%
\bibitem [{\citenamefont {Wang}\ \emph {et~al.}(2015)\citenamefont {Wang}, \citenamefont {Agterberg},\ and\ \citenamefont {Chubukov}}]{Wang_PRL2015}%
  \BibitemOpen
  \bibfield  {author} {\bibinfo {author} {\bibfnamefont {Y.}~\bibnamefont {Wang}}, \bibinfo {author} {\bibfnamefont {D.~F.}\ \bibnamefont {Agterberg}},\ and\ \bibinfo {author} {\bibfnamefont {A.}~\bibnamefont {Chubukov}},\ }\bibfield  {title} {\bibinfo {title} {Coexistence of {C}harge-{D}ensity-{W}ave and {P}air-{D}ensity-{W}ave {O}rders in {U}nderdoped {C}uprates},\ }\href {https://doi.org/10.1103/PhysRevLett.114.197001} {\bibfield  {journal} {\bibinfo  {journal} {Phys. Rev. Lett.}\ }\textbf {\bibinfo {volume} {114}},\ \bibinfo {pages} {197001} (\bibinfo {year} {2015})}\BibitemShut {NoStop}%
\bibitem [{\citenamefont {Wu}\ \emph {et~al.}(2023)\citenamefont {Wu}, \citenamefont {Nosov}, \citenamefont {Patel},\ and\ \citenamefont {Raghu}}]{Wu_PRL2023}%
  \BibitemOpen
  \bibfield  {author} {\bibinfo {author} {\bibfnamefont {Y.-M.}\ \bibnamefont {Wu}}, \bibinfo {author} {\bibfnamefont {P.~A.}\ \bibnamefont {Nosov}}, \bibinfo {author} {\bibfnamefont {A.~A.}\ \bibnamefont {Patel}},\ and\ \bibinfo {author} {\bibfnamefont {S.}~\bibnamefont {Raghu}},\ }\bibfield  {title} {\bibinfo {title} {Pair {D}ensity {W}ave {O}rder from {E}lectron {R}epulsion},\ }\href {https://doi.org/10.1103/PhysRevLett.130.026001} {\bibfield  {journal} {\bibinfo  {journal} {Phys. Rev. Lett.}\ }\textbf {\bibinfo {volume} {130}},\ \bibinfo {pages} {026001} (\bibinfo {year} {2023})}\BibitemShut {NoStop}%
\bibitem [{\citenamefont {Soto-Garrido}\ and\ \citenamefont {Fradkin}(2014)}]{Fradkin_PRB2014}%
  \BibitemOpen
  \bibfield  {author} {\bibinfo {author} {\bibfnamefont {R.}~\bibnamefont {Soto-Garrido}}\ and\ \bibinfo {author} {\bibfnamefont {E.}~\bibnamefont {Fradkin}},\ }\bibfield  {title} {\bibinfo {title} {Pair-density-wave superconducting states and electronic liquid-crystal phases},\ }\href {https://doi.org/10.1103/PhysRevB.89.165126} {\bibfield  {journal} {\bibinfo  {journal} {Phys. Rev. B}\ }\textbf {\bibinfo {volume} {89}},\ \bibinfo {pages} {165126} (\bibinfo {year} {2014})}\BibitemShut {NoStop}%
\bibitem [{\citenamefont {Ticea}\ \emph {et~al.}(2024)\citenamefont {Ticea}, \citenamefont {Raghu},\ and\ \citenamefont {Wu}}]{ticea_2024}%
  \BibitemOpen
  \bibfield  {author} {\bibinfo {author} {\bibfnamefont {N.~S.}\ \bibnamefont {Ticea}}, \bibinfo {author} {\bibfnamefont {S.}~\bibnamefont {Raghu}},\ and\ \bibinfo {author} {\bibfnamefont {Y.-M.}\ \bibnamefont {Wu}},\ }\href@noop {} {\bibinfo {title} {Pair density wave order in multiband systems}} (\bibinfo {year} {2024}),\ \Eprint {https://arxiv.org/abs/2403.00156} {arXiv:2403.00156 [cond-mat.supr-con]} \BibitemShut {NoStop}%
\bibitem [{\citenamefont {Yerin}\ \emph {et~al.}(2023)\citenamefont {Yerin}, \citenamefont {Drechsler}, \citenamefont {Cuoco},\ and\ \citenamefont {Petrillo}}]{Yerin_JPCM2023}%
  \BibitemOpen
  \bibfield  {author} {\bibinfo {author} {\bibfnamefont {Y.}~\bibnamefont {Yerin}}, \bibinfo {author} {\bibfnamefont {S.-L.}\ \bibnamefont {Drechsler}}, \bibinfo {author} {\bibfnamefont {M.}~\bibnamefont {Cuoco}},\ and\ \bibinfo {author} {\bibfnamefont {C.}~\bibnamefont {Petrillo}},\ }\bibfield  {title} {\bibinfo {title} {Multiple-q current states in a multicomponent superconducting channel},\ }\href {https://doi.org/10.1088/1361-648X/acf42d} {\bibfield  {journal} {\bibinfo  {journal} {Journal of Physics: Condensed Matter}\ }\textbf {\bibinfo {volume} {35}},\ \bibinfo {pages} {505601} (\bibinfo {year} {2023})}\BibitemShut {NoStop}%
\bibitem [{\citenamefont {Ng}\ and\ \citenamefont {Sigrist}(2000)}]{Ng2000EPL}%
  \BibitemOpen
  \bibfield  {author} {\bibinfo {author} {\bibfnamefont {K.~K.}\ \bibnamefont {Ng}}\ and\ \bibinfo {author} {\bibfnamefont {M.}~\bibnamefont {Sigrist}},\ }\bibfield  {title} {\bibinfo {title} {{The role of spin-orbit coupling for the superconducting state in {${\mathrm{Sr}}_{2}{\mathrm{RuO}}_{4}$} }},\ }\href {https://doi.org/10.1209/epl/i2000-00173-x} {\bibfield  {journal} {\bibinfo  {journal} {Europhys. Lett.}\ }\textbf {\bibinfo {volume} {49}},\ \bibinfo {pages} {473} (\bibinfo {year} {2000})}\BibitemShut {NoStop}%
\bibitem [{\citenamefont {Haverkort}\ \emph {et~al.}(2008)\citenamefont {Haverkort}, \citenamefont {Elfimov}, \citenamefont {Tjeng}, \citenamefont {Sawatzky},\ and\ \citenamefont {Damascelli}}]{Haverkort2008PRL}%
  \BibitemOpen
  \bibfield  {author} {\bibinfo {author} {\bibfnamefont {M.~W.}\ \bibnamefont {Haverkort}}, \bibinfo {author} {\bibfnamefont {I.~S.}\ \bibnamefont {Elfimov}}, \bibinfo {author} {\bibfnamefont {L.~H.}\ \bibnamefont {Tjeng}}, \bibinfo {author} {\bibfnamefont {G.~A.}\ \bibnamefont {Sawatzky}},\ and\ \bibinfo {author} {\bibfnamefont {A.}~\bibnamefont {Damascelli}},\ }\bibfield  {title} {\bibinfo {title} {Strong {S}pin-{O}rbit {C}oupling {E}ffects on the {F}ermi {S}urface of {${\mathrm{Sr}}_{2}{\mathrm{RuO}}_{4}$} and {${\mathrm{Sr}}_{2}{\mathrm{RhO}}_{4}$}},\ }\href {https://doi.org/10.1103/PhysRevLett.101.026406} {\bibfield  {journal} {\bibinfo  {journal} {Phys. Rev. Lett.}\ }\textbf {\bibinfo {volume} {101}},\ \bibinfo {pages} {026406} (\bibinfo {year} {2008})}\BibitemShut {NoStop}%
\bibitem [{\citenamefont {Kim}\ \emph {et~al.}(2018)\citenamefont {Kim}, \citenamefont {Mravlje}, \citenamefont {Ferrero}, \citenamefont {Parcollet},\ and\ \citenamefont {Georges}}]{Kim2018PRL}%
  \BibitemOpen
  \bibfield  {author} {\bibinfo {author} {\bibfnamefont {M.}~\bibnamefont {Kim}}, \bibinfo {author} {\bibfnamefont {J.}~\bibnamefont {Mravlje}}, \bibinfo {author} {\bibfnamefont {M.}~\bibnamefont {Ferrero}}, \bibinfo {author} {\bibfnamefont {O.}~\bibnamefont {Parcollet}},\ and\ \bibinfo {author} {\bibfnamefont {A.}~\bibnamefont {Georges}},\ }\bibfield  {title} {\bibinfo {title} {Spin-{O}rbit {C}oupling and {E}lectronic {C}orrelations in {${\mathrm{Sr}}_{2}{\mathrm{RuO}}_{4}$}},\ }\href {https://doi.org/10.1103/PhysRevLett.120.126401} {\bibfield  {journal} {\bibinfo  {journal} {Phys. Rev. Lett.}\ }\textbf {\bibinfo {volume} {120}},\ \bibinfo {pages} {126401} (\bibinfo {year} {2018})}\BibitemShut {NoStop}%
\bibitem [{\citenamefont {Tamai}\ \emph {et~al.}(2019)\citenamefont {Tamai}, \citenamefont {Zingl}, \citenamefont {Rozbicki}, \citenamefont {Cappelli}, \citenamefont {Ricc\`o}, \citenamefont {de~la Torre}, \citenamefont {McKeown~Walker}, \citenamefont {Bruno}, \citenamefont {King}, \citenamefont {Meevasana}, \citenamefont {Shi}, \citenamefont {Radovi\ifmmode~\acute{c}\else \'{c}\fi{}}, \citenamefont {Plumb}, \citenamefont {Gibbs}, \citenamefont {Mackenzie}, \citenamefont {Berthod}, \citenamefont {Strand}, \citenamefont {Kim}, \citenamefont {Georges},\ and\ \citenamefont {Baumberger}}]{Tamai2019PRX}%
  \BibitemOpen
  \bibfield  {author} {\bibinfo {author} {\bibfnamefont {A.}~\bibnamefont {Tamai}}, \bibinfo {author} {\bibfnamefont {M.}~\bibnamefont {Zingl}}, \bibinfo {author} {\bibfnamefont {E.}~\bibnamefont {Rozbicki}}, \bibinfo {author} {\bibfnamefont {E.}~\bibnamefont {Cappelli}}, \bibinfo {author} {\bibfnamefont {S.}~\bibnamefont {Ricc\`o}}, \bibinfo {author} {\bibfnamefont {A.}~\bibnamefont {de~la Torre}}, \bibinfo {author} {\bibfnamefont {S.}~\bibnamefont {McKeown~Walker}}, \bibinfo {author} {\bibfnamefont {F.~Y.}\ \bibnamefont {Bruno}}, \bibinfo {author} {\bibfnamefont {P.~D.~C.}\ \bibnamefont {King}}, \bibinfo {author} {\bibfnamefont {W.}~\bibnamefont {Meevasana}}, \bibinfo {author} {\bibfnamefont {M.}~\bibnamefont {Shi}}, \bibinfo {author} {\bibfnamefont {M.}~\bibnamefont {Radovi\ifmmode~\acute{c}\else \'{c}\fi{}}}, \bibinfo {author} {\bibfnamefont {N.~C.}\ \bibnamefont {Plumb}}, \bibinfo {author} {\bibfnamefont {A.~S.}\ \bibnamefont {Gibbs}}, \bibinfo {author} {\bibfnamefont {A.~P.}\ \bibnamefont {Mackenzie}},
  \bibinfo {author} {\bibfnamefont {C.}~\bibnamefont {Berthod}}, \bibinfo {author} {\bibfnamefont {H.~U.~R.}\ \bibnamefont {Strand}}, \bibinfo {author} {\bibfnamefont {M.}~\bibnamefont {Kim}}, \bibinfo {author} {\bibfnamefont {A.}~\bibnamefont {Georges}},\ and\ \bibinfo {author} {\bibfnamefont {F.}~\bibnamefont {Baumberger}},\ }\bibfield  {title} {\bibinfo {title} {High-{R}esolution {P}hotoemission on {${\mathrm{Sr}}_{2}{\mathrm{RuO}}_{4}$} {R}eveals {C}orrelation-{E}nhanced {E}ffective {S}pin-{O}rbit {C}oupling and {D}ominantly {L}ocal {S}elf-{E}nergies},\ }\href {https://doi.org/10.1103/PhysRevX.9.021048} {\bibfield  {journal} {\bibinfo  {journal} {Phys. Rev. X}\ }\textbf {\bibinfo {volume} {9}},\ \bibinfo {pages} {021048} (\bibinfo {year} {2019})}\BibitemShut {NoStop}%
\bibitem [{\citenamefont {Veenstra}\ \emph {et~al.}(2014)\citenamefont {Veenstra}, \citenamefont {Zhu}, \citenamefont {Raichle}, \citenamefont {Ludbrook}, \citenamefont {Nicolaou}, \citenamefont {Slomski}, \citenamefont {Landolt}, \citenamefont {Kittaka}, \citenamefont {Maeno}, \citenamefont {Dil}, \citenamefont {Elfimov}, \citenamefont {Haverkort},\ and\ \citenamefont {Damascelli}}]{Veenstra2014PRL}%
  \BibitemOpen
  \bibfield  {author} {\bibinfo {author} {\bibfnamefont {C.~N.}\ \bibnamefont {Veenstra}}, \bibinfo {author} {\bibfnamefont {Z.-H.}\ \bibnamefont {Zhu}}, \bibinfo {author} {\bibfnamefont {M.}~\bibnamefont {Raichle}}, \bibinfo {author} {\bibfnamefont {B.~M.}\ \bibnamefont {Ludbrook}}, \bibinfo {author} {\bibfnamefont {A.}~\bibnamefont {Nicolaou}}, \bibinfo {author} {\bibfnamefont {B.}~\bibnamefont {Slomski}}, \bibinfo {author} {\bibfnamefont {G.}~\bibnamefont {Landolt}}, \bibinfo {author} {\bibfnamefont {S.}~\bibnamefont {Kittaka}}, \bibinfo {author} {\bibfnamefont {Y.}~\bibnamefont {Maeno}}, \bibinfo {author} {\bibfnamefont {J.~H.}\ \bibnamefont {Dil}}, \bibinfo {author} {\bibfnamefont {I.~S.}\ \bibnamefont {Elfimov}}, \bibinfo {author} {\bibfnamefont {M.~W.}\ \bibnamefont {Haverkort}},\ and\ \bibinfo {author} {\bibfnamefont {A.}~\bibnamefont {Damascelli}},\ }\bibfield  {title} {\bibinfo {title} {Spin-{O}rbital {E}ntanglement and the {B}reakdown of {S}inglets and {T}riplets in
  {${\mathrm{Sr}}_{2}{\mathrm{RuO}}_{4}$} {R}evealed by {S}pin- and {A}ngle-{R}esolved {P}hotoemission {S}pectroscopy},\ }\href {https://doi.org/10.1103/PhysRevLett.112.127002} {\bibfield  {journal} {\bibinfo  {journal} {Phys. Rev. Lett.}\ }\textbf {\bibinfo {volume} {112}},\ \bibinfo {pages} {127002} (\bibinfo {year} {2014})}\BibitemShut {NoStop}%
\bibitem [{\citenamefont {Mravlje}\ \emph {et~al.}(2011)\citenamefont {Mravlje}, \citenamefont {Aichhorn}, \citenamefont {Miyake}, \citenamefont {Haule}, \citenamefont {Kotliar},\ and\ \citenamefont {Georges}}]{mravlje2011PRL}%
  \BibitemOpen
  \bibfield  {author} {\bibinfo {author} {\bibfnamefont {J.}~\bibnamefont {Mravlje}}, \bibinfo {author} {\bibfnamefont {M.}~\bibnamefont {Aichhorn}}, \bibinfo {author} {\bibfnamefont {T.}~\bibnamefont {Miyake}}, \bibinfo {author} {\bibfnamefont {K.}~\bibnamefont {Haule}}, \bibinfo {author} {\bibfnamefont {G.}~\bibnamefont {Kotliar}},\ and\ \bibinfo {author} {\bibfnamefont {A.}~\bibnamefont {Georges}},\ }\bibfield  {title} {\bibinfo {title} {Coherence-{I}ncoherence {C}rossover and the {M}ass-{R}enormalization {P}uzzles in {${\mathrm{Sr}}_{2}{\mathrm{RuO}}_{4}$}},\ }\href {https://link.aps.org/doi/10.1103/PhysRevLett.106.096401} {\bibfield  {journal} {\bibinfo  {journal} {Phys. Rev. Lett.}\ }\textbf {\bibinfo {volume} {106}},\ \bibinfo {pages} {096401} (\bibinfo {year} {2011})}\BibitemShut {NoStop}%
\bibitem [{\citenamefont {Georges}\ \emph {et~al.}(2013)\citenamefont {Georges}, \citenamefont {Medici},\ and\ \citenamefont {Mravlje}}]{Georges2013ARCMP}%
  \BibitemOpen
  \bibfield  {author} {\bibinfo {author} {\bibfnamefont {A.}~\bibnamefont {Georges}}, \bibinfo {author} {\bibfnamefont {L.~d.}\ \bibnamefont {Medici}},\ and\ \bibinfo {author} {\bibfnamefont {J.}~\bibnamefont {Mravlje}},\ }\bibfield  {title} {\bibinfo {title} {Strong {C}orrelations from {H}und’s {C}oupling},\ }\href {https://doi.org/10.1146/annurev-conmatphys-020911-125045} {\bibfield  {journal} {\bibinfo  {journal} {Annu. Rev. Condens. Matter Phys.}\ }\textbf {\bibinfo {volume} {4}},\ \bibinfo {pages} {137} (\bibinfo {year} {2013})}\BibitemShut {NoStop}%
\bibitem [{\citenamefont {Klejnberg}\ and\ \citenamefont {Spalek}(1999)}]{klejnberg1999hund}%
  \BibitemOpen
  \bibfield  {author} {\bibinfo {author} {\bibfnamefont {A.}~\bibnamefont {Klejnberg}}\ and\ \bibinfo {author} {\bibfnamefont {J.}~\bibnamefont {Spalek}},\ }\bibfield  {title} {\bibinfo {title} {Hund's rule coupling as the microscopic origin of the spin-triplet pairing in a correlated and degenerate band system},\ }\href {https://doi.org/10.1088/0953-8984/11/34/307} {\bibfield  {journal} {\bibinfo  {journal} {J. Phys.: Condens. Matter}\ }\textbf {\bibinfo {volume} {11}},\ \bibinfo {pages} {6553} (\bibinfo {year} {1999})}\BibitemShut {NoStop}%
\bibitem [{\citenamefont {Han}(2004)}]{HanPRB2004}%
  \BibitemOpen
  \bibfield  {author} {\bibinfo {author} {\bibfnamefont {J.~E.}\ \bibnamefont {Han}},\ }\bibfield  {title} {\bibinfo {title} {Spin-triplet $s$-wave local pairing induced by {H}und's rule coupling},\ }\href {https://doi.org/10.1103/PhysRevB.70.054513} {\bibfield  {journal} {\bibinfo  {journal} {Phys. Rev. B}\ }\textbf {\bibinfo {volume} {70}},\ \bibinfo {pages} {054513} (\bibinfo {year} {2004})}\BibitemShut {NoStop}%
\bibitem [{\citenamefont {Ramires}\ and\ \citenamefont {Sigrist}(2019)}]{Ramires2019PRB}%
  \BibitemOpen
  \bibfield  {author} {\bibinfo {author} {\bibfnamefont {A.}~\bibnamefont {Ramires}}\ and\ \bibinfo {author} {\bibfnamefont {M.}~\bibnamefont {Sigrist}},\ }\bibfield  {title} {\bibinfo {title} {Superconducting order parameter of {${\mathrm{Sr}}_{2}{\mathrm{RuO}}_{4}$}: A microscopic perspective},\ }\href {https://doi.org/10.1103/PhysRevB.100.104501} {\bibfield  {journal} {\bibinfo  {journal} {Phys. Rev. B}\ }\textbf {\bibinfo {volume} {100}},\ \bibinfo {pages} {104501} (\bibinfo {year} {2019})}\BibitemShut {NoStop}%
\bibitem [{\citenamefont {Kaba}\ and\ \citenamefont {S\'en\'echal}(2019)}]{kaba_PRB2019}%
  \BibitemOpen
  \bibfield  {author} {\bibinfo {author} {\bibfnamefont {S.-O.}\ \bibnamefont {Kaba}}\ and\ \bibinfo {author} {\bibfnamefont {D.}~\bibnamefont {S\'en\'echal}},\ }\bibfield  {title} {\bibinfo {title} {Group-theoretical classification of superconducting states of strontium ruthenate},\ }\href {https://doi.org/10.1103/PhysRevB.100.214507} {\bibfield  {journal} {\bibinfo  {journal} {Phys. Rev. B}\ }\textbf {\bibinfo {volume} {100}},\ \bibinfo {pages} {214507} (\bibinfo {year} {2019})}\BibitemShut {NoStop}%
\bibitem [{\citenamefont {Cheung}\ and\ \citenamefont {Agterberg}(2019)}]{Cheung2019PRB}%
  \BibitemOpen
  \bibfield  {author} {\bibinfo {author} {\bibfnamefont {A.~K.~C.}\ \bibnamefont {Cheung}}\ and\ \bibinfo {author} {\bibfnamefont {D.~F.}\ \bibnamefont {Agterberg}},\ }\bibfield  {title} {\bibinfo {title} {Superconductivity in the presence of spin-orbit interactions stabilized by {H}und coupling},\ }\href {https://doi.org/10.1103/PhysRevB.99.024516} {\bibfield  {journal} {\bibinfo  {journal} {Phys. Rev. B}\ }\textbf {\bibinfo {volume} {99}},\ \bibinfo {pages} {024516} (\bibinfo {year} {2019})}\BibitemShut {NoStop}%
\bibitem [{\citenamefont {Clepkens}\ \emph {et~al.}(2021{\natexlab{b}})\citenamefont {Clepkens}, \citenamefont {Lindquist},\ and\ \citenamefont {Kee}}]{clepkens2021}%
  \BibitemOpen
  \bibfield  {author} {\bibinfo {author} {\bibfnamefont {J.}~\bibnamefont {Clepkens}}, \bibinfo {author} {\bibfnamefont {A.~W.}\ \bibnamefont {Lindquist}},\ and\ \bibinfo {author} {\bibfnamefont {H.-Y.}\ \bibnamefont {Kee}},\ }\bibfield  {title} {\bibinfo {title} {Shadowed triplet pairings in {H}und's metals with spin-orbit coupling},\ }\href {https://journals.aps.org/prresearch/abstract/10.1103/PhysRevResearch.3.013001} {\bibfield  {journal} {\bibinfo  {journal} {Phys. Rev. Res.}\ }\textbf {\bibinfo {volume} {3}},\ \bibinfo {pages} {013001} (\bibinfo {year} {2021}{\natexlab{b}})}\BibitemShut {NoStop}%
\bibitem [{\citenamefont {Salamone}\ \emph {et~al.}(2023)\citenamefont {Salamone}, \citenamefont {Hugdal}, \citenamefont {Jacobsen},\ and\ \citenamefont {Amundsen}}]{salamone_PRB2023}%
  \BibitemOpen
  \bibfield  {author} {\bibinfo {author} {\bibfnamefont {T.}~\bibnamefont {Salamone}}, \bibinfo {author} {\bibfnamefont {H.~G.}\ \bibnamefont {Hugdal}}, \bibinfo {author} {\bibfnamefont {S.~H.}\ \bibnamefont {Jacobsen}},\ and\ \bibinfo {author} {\bibfnamefont {M.}~\bibnamefont {Amundsen}},\ }\bibfield  {title} {\bibinfo {title} {High magnetic field superconductivity in a two-band superconductor},\ }\href {https://journals.aps.org/prb/abstract/10.1103/PhysRevB.107.174516} {\bibfield  {journal} {\bibinfo  {journal} {Phys. Rev. B}\ }\textbf {\bibinfo {volume} {107}},\ \bibinfo {pages} {174516} (\bibinfo {year} {2023})}\BibitemShut {NoStop}%
\bibitem [{\citenamefont {Daghofer}\ \emph {et~al.}(2010)\citenamefont {Daghofer}, \citenamefont {Nicholson}, \citenamefont {Moreo},\ and\ \citenamefont {Dagotto}}]{daghofer_PRB2010}%
  \BibitemOpen
  \bibfield  {author} {\bibinfo {author} {\bibfnamefont {M.}~\bibnamefont {Daghofer}}, \bibinfo {author} {\bibfnamefont {A.}~\bibnamefont {Nicholson}}, \bibinfo {author} {\bibfnamefont {A.}~\bibnamefont {Moreo}},\ and\ \bibinfo {author} {\bibfnamefont {E.}~\bibnamefont {Dagotto}},\ }\bibfield  {title} {\bibinfo {title} {Three orbital model for the iron-based superconductors},\ }\href {https://journals.aps.org/prb/abstract/10.1103/PhysRevB.81.014511} {\bibfield  {journal} {\bibinfo  {journal} {Phys. Rev. B}\ }\textbf {\bibinfo {volume} {81}},\ \bibinfo {pages} {014511} (\bibinfo {year} {2010})}\BibitemShut {NoStop}%
\bibitem [{\citenamefont {Maeno}\ \emph {et~al.}(2024)\citenamefont {Maeno}, \citenamefont {Yonezawa},\ and\ \citenamefont {Ramires}}]{maeno_2024}%
  \BibitemOpen
  \bibfield  {author} {\bibinfo {author} {\bibfnamefont {Y.}~\bibnamefont {Maeno}}, \bibinfo {author} {\bibfnamefont {S.}~\bibnamefont {Yonezawa}},\ and\ \bibinfo {author} {\bibfnamefont {A.}~\bibnamefont {Ramires}},\ }\bibfield  {title} {\bibinfo {title} {Still mystery after all these years—{U}nconventional superconductivity of {${\mathrm{Sr}}_{2}{\mathrm{RuO}}_{4}$}—},\ }\href {https://journals.jps.jp/doi/full/10.7566/JPSJ.93.062001?mobileUi=0} {\bibfield  {journal} {\bibinfo  {journal} {J. Phys. Soc. Jpn.}\ }\textbf {\bibinfo {volume} {93}},\ \bibinfo {pages} {062001} (\bibinfo {year} {2024})}\BibitemShut {NoStop}%
\bibitem [{\citenamefont {Inaba}\ and\ \citenamefont {Suga}(2012)}]{Inaba_PRL2012}%
  \BibitemOpen
  \bibfield  {author} {\bibinfo {author} {\bibfnamefont {K.}~\bibnamefont {Inaba}}\ and\ \bibinfo {author} {\bibfnamefont {S.-i.}\ \bibnamefont {Suga}},\ }\bibfield  {title} {\bibinfo {title} {Superfluid {S}tate of {R}epulsively {I}nteracting {T}hree-{C}omponent {F}ermionic {A}toms in {O}ptical {L}attices},\ }\href {https://doi.org/10.1103/PhysRevLett.108.255301} {\bibfield  {journal} {\bibinfo  {journal} {Phys. Rev. Lett.}\ }\textbf {\bibinfo {volume} {108}},\ \bibinfo {pages} {255301} (\bibinfo {year} {2012})}\BibitemShut {NoStop}%
\bibitem [{\citenamefont {Shimahara}\ and\ \citenamefont {Rainer}(1997)}]{shimahara_JPSJ1997}%
  \BibitemOpen
  \bibfield  {author} {\bibinfo {author} {\bibfnamefont {H.}~\bibnamefont {Shimahara}}\ and\ \bibinfo {author} {\bibfnamefont {D.}~\bibnamefont {Rainer}},\ }\bibfield  {title} {\bibinfo {title} {Crossover from {V}ortex {S}tates to the {F}ulde-{F}errell-{L}arkin-{O}vchinnikov {S}tate in {T}wo-{D}imensional s-and d-{W}ave {S}uperconductors},\ }\href {https://journals.jps.jp/doi/10.1143/JPSJ.66.3591} {\bibfield  {journal} {\bibinfo  {journal} {J. Phys. Soc. Jpn.}\ }\textbf {\bibinfo {volume} {66}},\ \bibinfo {pages} {3591} (\bibinfo {year} {1997})}\BibitemShut {NoStop}%
\bibitem [{\citenamefont {Yang}\ and\ \citenamefont {Sondhi}(1998)}]{Yang_PRB1998}%
  \BibitemOpen
  \bibfield  {author} {\bibinfo {author} {\bibfnamefont {K.}~\bibnamefont {Yang}}\ and\ \bibinfo {author} {\bibfnamefont {S.~L.}\ \bibnamefont {Sondhi}},\ }\bibfield  {title} {\bibinfo {title} {Response of a ${d}_{{x}^{2}\ensuremath{-}{y}^{2}}$ superconductor to a {Z}eeman magnetic field},\ }\href {https://doi.org/10.1103/PhysRevB.57.8566} {\bibfield  {journal} {\bibinfo  {journal} {Phys. Rev. B}\ }\textbf {\bibinfo {volume} {57}},\ \bibinfo {pages} {8566} (\bibinfo {year} {1998})}\BibitemShut {NoStop}%
\bibitem [{\citenamefont {Aoki}\ \emph {et~al.}(2022)\citenamefont {Aoki}, \citenamefont {Brison}, \citenamefont {Flouquet}, \citenamefont {Ishida}, \citenamefont {Knebel}, \citenamefont {Tokunaga},\ and\ \citenamefont {Yanase}}]{aoki_JPCM2022}%
  \BibitemOpen
  \bibfield  {author} {\bibinfo {author} {\bibfnamefont {D.}~\bibnamefont {Aoki}}, \bibinfo {author} {\bibfnamefont {J.-P.}\ \bibnamefont {Brison}}, \bibinfo {author} {\bibfnamefont {J.}~\bibnamefont {Flouquet}}, \bibinfo {author} {\bibfnamefont {K.}~\bibnamefont {Ishida}}, \bibinfo {author} {\bibfnamefont {G.}~\bibnamefont {Knebel}}, \bibinfo {author} {\bibfnamefont {Y.}~\bibnamefont {Tokunaga}},\ and\ \bibinfo {author} {\bibfnamefont {Y.}~\bibnamefont {Yanase}},\ }\bibfield  {title} {\bibinfo {title} {Unconventional superconductivity in {$\mathrm{UTe}_{2}$}},\ }\href {https://iopscience.iop.org/article/10.1088/1361-648X/ac5863} {\bibfield  {journal} {\bibinfo  {journal} {J. Phys.: Condens. Matter}\ }\textbf {\bibinfo {volume} {34}},\ \bibinfo {pages} {243002} (\bibinfo {year} {2022})}\BibitemShut {NoStop}%
\bibitem [{\citenamefont {Borisenko}\ \emph {et~al.}(2016)\citenamefont {Borisenko}, \citenamefont {Evtushinsky}, \citenamefont {Liu}, \citenamefont {Morozov}, \citenamefont {Kappenberger}, \citenamefont {Wurmehl}, \citenamefont {B{\"u}chner}, \citenamefont {Yaresko}, \citenamefont {Kim}, \citenamefont {Hoesch} \emph {et~al.}}]{borisenko_Nat2016}%
  \BibitemOpen
  \bibfield  {author} {\bibinfo {author} {\bibfnamefont {S.}~\bibnamefont {Borisenko}}, \bibinfo {author} {\bibfnamefont {D.}~\bibnamefont {Evtushinsky}}, \bibinfo {author} {\bibfnamefont {Z.-H.}\ \bibnamefont {Liu}}, \bibinfo {author} {\bibfnamefont {I.}~\bibnamefont {Morozov}}, \bibinfo {author} {\bibfnamefont {R.}~\bibnamefont {Kappenberger}}, \bibinfo {author} {\bibfnamefont {S.}~\bibnamefont {Wurmehl}}, \bibinfo {author} {\bibfnamefont {B.}~\bibnamefont {B{\"u}chner}}, \bibinfo {author} {\bibfnamefont {A.}~\bibnamefont {Yaresko}}, \bibinfo {author} {\bibfnamefont {T.}~\bibnamefont {Kim}}, \bibinfo {author} {\bibfnamefont {M.}~\bibnamefont {Hoesch}}, \emph {et~al.},\ }\bibfield  {title} {\bibinfo {title} {Direct observation of spin--orbit coupling in iron-based superconductors},\ }\href {https://www.nature.com/articles/nphys3594} {\bibfield  {journal} {\bibinfo  {journal} {Nat. Phys.}\ }\textbf {\bibinfo {volume} {12}},\ \bibinfo {pages} {311} (\bibinfo {year} {2016})}\BibitemShut {NoStop}%
\bibitem [{\citenamefont {Zegrodnik}\ and\ \citenamefont {Spa{\l}ek}(2015)}]{zegrodnik_JSNM2015}%
  \BibitemOpen
  \bibfield  {author} {\bibinfo {author} {\bibfnamefont {M.}~\bibnamefont {Zegrodnik}}\ and\ \bibinfo {author} {\bibfnamefont {J.}~\bibnamefont {Spa{\l}ek}},\ }\bibfield  {title} {\bibinfo {title} {Spontaneous {A}ppearance of the {S}pin-{T}riplet {F}ulde-{F}errell-{L}arkin-{O}vchinnikov {P}hase in a {T}wo-{B}and {M}odel: {P}ossible {A}pplication to {$\mathrm{LaFeAsO}_{1-x}\mathrm{F}_{x}$}},\ }\href {https://link.springer.com/article/10.1007/s10948-014-2800-0} {\bibfield  {journal} {\bibinfo  {journal} {J. Supercond. Novel Magn.}\ }\textbf {\bibinfo {volume} {28}},\ \bibinfo {pages} {1155} (\bibinfo {year} {2015})}\BibitemShut {NoStop}%
\bibitem [{\citenamefont {Black-Schaffer}\ and\ \citenamefont {Balatsky}(2013)}]{AnnicaPRB2013}%
  \BibitemOpen
  \bibfield  {author} {\bibinfo {author} {\bibfnamefont {A.~M.}\ \bibnamefont {Black-Schaffer}}\ and\ \bibinfo {author} {\bibfnamefont {A.~V.}\ \bibnamefont {Balatsky}},\ }\bibfield  {title} {\bibinfo {title} {Odd-frequency superconducting pairing in multiband superconductors},\ }\href {https://doi.org/10.1103/PhysRevB.88.104514} {\bibfield  {journal} {\bibinfo  {journal} {Phys. Rev. B}\ }\textbf {\bibinfo {volume} {88}},\ \bibinfo {pages} {104514} (\bibinfo {year} {2013})}\BibitemShut {NoStop}%
\bibitem [{\citenamefont {Komendov\'a}\ and\ \citenamefont {Black-Schaffer}(2017)}]{Komendova_PRL2017}%
  \BibitemOpen
  \bibfield  {author} {\bibinfo {author} {\bibfnamefont {L.}~\bibnamefont {Komendov\'a}}\ and\ \bibinfo {author} {\bibfnamefont {A.~M.}\ \bibnamefont {Black-Schaffer}},\ }\bibfield  {title} {\bibinfo {title} {Odd-{F}requency {S}uperconductivity in {${\mathrm{Sr}}_{2}{\mathrm{RuO}}_{4}$} {M}easured by {K}err {R}otation},\ }\href {https://doi.org/10.1103/PhysRevLett.119.087001} {\bibfield  {journal} {\bibinfo  {journal} {Phys. Rev. Lett.}\ }\textbf {\bibinfo {volume} {119}},\ \bibinfo {pages} {087001} (\bibinfo {year} {2017})}\BibitemShut {NoStop}%
\bibitem [{\citenamefont {Chakraborty}\ and\ \citenamefont {Black-Schaffer}(2022)}]{Chakraborty_PRB2022}%
  \BibitemOpen
  \bibfield  {author} {\bibinfo {author} {\bibfnamefont {D.}~\bibnamefont {Chakraborty}}\ and\ \bibinfo {author} {\bibfnamefont {A.~M.}\ \bibnamefont {Black-Schaffer}},\ }\bibfield  {title} {\bibinfo {title} {Interplay of finite-energy and finite-momentum superconducting pairing},\ }\href {https://doi.org/10.1103/PhysRevB.106.024511} {\bibfield  {journal} {\bibinfo  {journal} {Phys. Rev. B}\ }\textbf {\bibinfo {volume} {106}},\ \bibinfo {pages} {024511} (\bibinfo {year} {2022})}\BibitemShut {NoStop}%
\bibitem [{\citenamefont {Fanfarillo}\ \emph {et~al.}(2020)\citenamefont {Fanfarillo}, \citenamefont {Valli},\ and\ \citenamefont {Capone}}]{Fanfarillo_PRL2020}%
  \BibitemOpen
  \bibfield  {author} {\bibinfo {author} {\bibfnamefont {L.}~\bibnamefont {Fanfarillo}}, \bibinfo {author} {\bibfnamefont {A.}~\bibnamefont {Valli}},\ and\ \bibinfo {author} {\bibfnamefont {M.}~\bibnamefont {Capone}},\ }\bibfield  {title} {\bibinfo {title} {Synergy between {H}und-{D}riven {C}orrelations and {B}oson-{M}ediated {S}uperconductivity},\ }\href {https://doi.org/10.1103/PhysRevLett.125.177001} {\bibfield  {journal} {\bibinfo  {journal} {Phys. Rev. Lett.}\ }\textbf {\bibinfo {volume} {125}},\ \bibinfo {pages} {177001} (\bibinfo {year} {2020})}\BibitemShut {NoStop}%
\bibitem [{\citenamefont {Fanfarillo}\ \emph {et~al.}(2023)\citenamefont {Fanfarillo}, \citenamefont {Valli},\ and\ \citenamefont {Capone}}]{Fanfarillo_PRB2023}%
  \BibitemOpen
  \bibfield  {author} {\bibinfo {author} {\bibfnamefont {L.}~\bibnamefont {Fanfarillo}}, \bibinfo {author} {\bibfnamefont {A.}~\bibnamefont {Valli}},\ and\ \bibinfo {author} {\bibfnamefont {M.}~\bibnamefont {Capone}},\ }\bibfield  {title} {\bibinfo {title} {Nematic spectral signatures of the {H}und's metal},\ }\href {https://doi.org/10.1103/PhysRevB.107.L081114} {\bibfield  {journal} {\bibinfo  {journal} {Phys. Rev. B}\ }\textbf {\bibinfo {volume} {107}},\ \bibinfo {pages} {L081114} (\bibinfo {year} {2023})}\BibitemShut {NoStop}%
\bibitem [{\citenamefont {Gingras}\ \emph {et~al.}(2022)\citenamefont {Gingras}, \citenamefont {Allaglo}, \citenamefont {Nourafkan}, \citenamefont {C\^ot\'e},\ and\ \citenamefont {Tremblay}}]{Gingras_PRB2022}%
  \BibitemOpen
  \bibfield  {author} {\bibinfo {author} {\bibfnamefont {O.}~\bibnamefont {Gingras}}, \bibinfo {author} {\bibfnamefont {N.}~\bibnamefont {Allaglo}}, \bibinfo {author} {\bibfnamefont {R.}~\bibnamefont {Nourafkan}}, \bibinfo {author} {\bibfnamefont {M.}~\bibnamefont {C\^ot\'e}},\ and\ \bibinfo {author} {\bibfnamefont {A.-M.~S.}\ \bibnamefont {Tremblay}},\ }\bibfield  {title} {\bibinfo {title} {Superconductivity in correlated multiorbital systems with spin-orbit coupling: Coexistence of even- and odd-frequency pairing, and the case of {${\mathrm{Sr}}_{2}{\mathrm{RuO}}_{4}$}},\ }\href {https://doi.org/10.1103/PhysRevB.106.064513} {\bibfield  {journal} {\bibinfo  {journal} {Phys. Rev. B}\ }\textbf {\bibinfo {volume} {106}},\ \bibinfo {pages} {064513} (\bibinfo {year} {2022})}\BibitemShut {NoStop}%
\bibitem [{\citenamefont {Kugler}\ \emph {et~al.}(2020)\citenamefont {Kugler}, \citenamefont {Zingl}, \citenamefont {Strand}, \citenamefont {Lee}, \citenamefont {von Delft},\ and\ \citenamefont {Georges}}]{Kugler_PRL2020}%
  \BibitemOpen
  \bibfield  {author} {\bibinfo {author} {\bibfnamefont {F.~B.}\ \bibnamefont {Kugler}}, \bibinfo {author} {\bibfnamefont {M.}~\bibnamefont {Zingl}}, \bibinfo {author} {\bibfnamefont {H.~U.~R.}\ \bibnamefont {Strand}}, \bibinfo {author} {\bibfnamefont {S.-S.~B.}\ \bibnamefont {Lee}}, \bibinfo {author} {\bibfnamefont {J.}~\bibnamefont {von Delft}},\ and\ \bibinfo {author} {\bibfnamefont {A.}~\bibnamefont {Georges}},\ }\bibfield  {title} {\bibinfo {title} {Strongly {C}orrelated {M}aterials from a {N}umerical {R}enormalization {G}roup {P}erspective: {H}ow the {F}ermi-{L}iquid {S}tate of {${\mathrm{Sr}}_{2}{\mathrm{RuO}}_{4}$} {E}merges},\ }\href {https://doi.org/10.1103/PhysRevLett.124.016401} {\bibfield  {journal} {\bibinfo  {journal} {Phys. Rev. Lett.}\ }\textbf {\bibinfo {volume} {124}},\ \bibinfo {pages} {016401} (\bibinfo {year} {2020})}\BibitemShut {NoStop}%
\bibitem [{\citenamefont {Blesio}\ \emph {et~al.}(2024)\citenamefont {Blesio}, \citenamefont {Beck}, \citenamefont {Gingras}, \citenamefont {Georges},\ and\ \citenamefont {Mravlje}}]{Blesio_PRR2024}%
  \BibitemOpen
  \bibfield  {author} {\bibinfo {author} {\bibfnamefont {G.}~\bibnamefont {Blesio}}, \bibinfo {author} {\bibfnamefont {S.}~\bibnamefont {Beck}}, \bibinfo {author} {\bibfnamefont {O.}~\bibnamefont {Gingras}}, \bibinfo {author} {\bibfnamefont {A.}~\bibnamefont {Georges}},\ and\ \bibinfo {author} {\bibfnamefont {J.}~\bibnamefont {Mravlje}},\ }\bibfield  {title} {\bibinfo {title} {Signatures of {H}und metal and finite-frequency nesting in {${\mathrm{Sr}}_{2}{\mathrm{RuO}}_{4}$} revealed by electronic {R}aman scattering},\ }\href {https://doi.org/10.1103/PhysRevResearch.6.023124} {\bibfield  {journal} {\bibinfo  {journal} {Phys. Rev. Res.}\ }\textbf {\bibinfo {volume} {6}},\ \bibinfo {pages} {023124} (\bibinfo {year} {2024})}\BibitemShut {NoStop}%
\bibitem [{\citenamefont {Lichtenstein}\ and\ \citenamefont {Katsnelson}(2000)}]{Lichtenstein_PRB2000}%
  \BibitemOpen
  \bibfield  {author} {\bibinfo {author} {\bibfnamefont {A.~I.}\ \bibnamefont {Lichtenstein}}\ and\ \bibinfo {author} {\bibfnamefont {M.~I.}\ \bibnamefont {Katsnelson}},\ }\bibfield  {title} {\bibinfo {title} {Antiferromagnetism and d-wave superconductivity in cuprates: {A} cluster dynamical mean-field theory},\ }\href {https://doi.org/10.1103/PhysRevB.62.R9283} {\bibfield  {journal} {\bibinfo  {journal} {Phys. Rev. B}\ }\textbf {\bibinfo {volume} {62}},\ \bibinfo {pages} {R9283} (\bibinfo {year} {2000})}\BibitemShut {NoStop}%
\bibitem [{\citenamefont {Kotliar}\ \emph {et~al.}(2001)\citenamefont {Kotliar}, \citenamefont {Savrasov}, \citenamefont {P\'alsson},\ and\ \citenamefont {Biroli}}]{Kotliar_PRL2001}%
  \BibitemOpen
  \bibfield  {author} {\bibinfo {author} {\bibfnamefont {G.}~\bibnamefont {Kotliar}}, \bibinfo {author} {\bibfnamefont {S.~Y.}\ \bibnamefont {Savrasov}}, \bibinfo {author} {\bibfnamefont {G.}~\bibnamefont {P\'alsson}},\ and\ \bibinfo {author} {\bibfnamefont {G.}~\bibnamefont {Biroli}},\ }\bibfield  {title} {\bibinfo {title} {Cellular {D}ynamical {M}ean {F}ield {A}pproach to {S}trongly {C}orrelated {S}ystems},\ }\href {https://doi.org/10.1103/PhysRevLett.87.186401} {\bibfield  {journal} {\bibinfo  {journal} {Phys. Rev. Lett.}\ }\textbf {\bibinfo {volume} {87}},\ \bibinfo {pages} {186401} (\bibinfo {year} {2001})}\BibitemShut {NoStop}%
\end{thebibliography}%

\end{document}